\documentclass[aps,prb,twocolumn]{revtex4-1}
\usepackage{graphicx}
\usepackage{amssymb}%
\usepackage{amsmath}
\usepackage{epsfig}
\usepackage{wasysym}
\usepackage{ulem}
\usepackage{color}
\usepackage{longtable}
\begin{document}
\title{Colorings of odd or even chirality on hexagonal lattices}
\author{O. C\'epas}
\affiliation{Institut N\'eel, CNRS, Universit\'e Grenoble Alpes, BP 166, F-38042 Grenoble cedex 9, France }

\begin{abstract}
We define two classes of colorings that have odd or even chirality on
hexagonal lattices.  This parity is an invariant in the dynamics of
all loops, and explains why standard Monte-Carlo algorithms are nonergodic.
We argue that adding the
motion of ``stranded'' loops allows for parity changes.  By
implementing this algorithm, we show that the even and odd classes
have the same entropy. In general, they do not have the same number of
states, except for the special geometry of long strips, where a Z$_2$
symmetry between even and odd states occurs in the thermodynamic
limit.
\end{abstract}

\maketitle
\section{Introduction}

Constructing classical states that satisfy some local constraints
everywhere on a lattice may be a difficult problem. The simplest physical
examples of such problems include dimer coverings of lattices, perfect
tilings of a surface with tiles of special geometry, coloring sites
with distinct colors, folding paper (origami). They have in common
that they can be formulated with ice-type or vertex models. The
calculation of the number of states and the thermodynamics is
difficult, but has been solved exactly in some cases. There are also interesting issues
regarding conservation laws in the dynamics and hidden higher symmetries.

A way to construct numerically such constrained states is to iterate a dynamical
process, starting from a simple state.  In particular, classical
Monte-Carlo algorithms sample the ensemble of states by flipping
variables collectively in order to preserve the constraints,\cite{Newman} either along
loops\cite{Stillinger} or in clusters of sites.\cite{Wang}  There are
now examples of models where such algorithms are nonergodic, and
conservation laws may hinder classes of states.  This is the case of
some dimer models on lattices in three spatial dimensions, where the
nonergodicity is restricted to flipping the smallest
loops.\cite{Sikora,Freedman} This is also generically the case when
periodic boundary conditions create sectors characterized by winding
numbers: here again the ergodicity is simply restored by including in
the dynamics the loops that wind accross the boundaries.  

A different example is given by the three-color Baxter's
model\cite{Baxter} that we will consider here. In this model, the
constraint consists of coloring the edges of a regular hexagonal
lattice with three colors such that no two neighbors have the same
color. Baxter calculated exactly the entropy of the three-colorings
for a hexagonal lattice with open boundary conditions. In order to
compute some observables, loop Monte-Carlo algorithms have been set up
but it has been recognized that they are nonergodic for lattices with
periodic boundary conditions,\cite{Huse,Mohar} possibly leading to
systematic errors. In particular, some states that are not connected
by the dynamics, have been constructed on finite-size systems, leading
to the conclusion that there are more than one class of
states.\cite{Mohar} Since this property is intimately related to the
topology of the two-dimensional lattice, an alternative Monte-Carlo
approach\cite{Moore} is to use other geometries for the boundaries,
such as a plane with open boundary conditions (but it suffers from
more finite-size effects) or a ``projective plane'', although the
hexagonal lattice has some noncubic vertices in this case.

Restricting to standard periodic boundary conditions and studying why
this dynamics in particular is nonergodic is nevertheless interesting, as a point of
principle.  This dynamics is not only used in Monte-Carlo algorithms
but also in quantum models constructed from these constrained states
(see Refs.~\onlinecite{Castelnovo,Cepas} in the coloring context).
Nonergodicity is generally related to some conservation
laws (or broken symmetries) which may have some consequences in
various problems.  It is also used in applications, \textit{e.g.} as
bits, and some proposals emphasized the advantage of the topological
nature of the ergodicity breaking in dimer models.\cite{Ioffe} Even
though imposing periodic boundary conditions seems to be an
``academic'' problem -they are not that of crystals-, it is, in
principle, possible to (i) design artificial superconducting
or magnetic devices with special geometry of the boundaries,\cite{realpbc} (ii)
consider molecular nanomagnets that realize special topologies,
\textit{e.g.} a sphere or a ring in the ``keplerate'' family.\cite{keplerate}  
Another close model is the three-coloring model with
an additional achiral constraint on the states which generates an infinite
number of sectors, even when winding loops are allowed to
flip.\cite{Fendley,Korshunov}

In the absence of ergodic algorithms to construct all states, simple
properties such as the number of missed states are not known. Is this
number extensive with the system size or not? If it is, what is the
entropy of the new class? Given a state, how do we know whether it is
connected to a simple reference state of a given class, \textit{i.e.}
what is the reason of the obstruction to ergodicity? Can we construct
an ergodic algorithm that will visit all states? We answer these
questions in the present paper.

The paper is organized as follows. We define the model in section
\ref{Model}, and we construct all three-coloring states on small
clusters in section \ref{enumeration}, from dimer coverings. We check
the numbers of three-colorings by using the method of transfer
matrix in section \ref{Tmc}. In section \ref{dyn}, we study the
dynamics of loops. In addition to standard winding-number classes
which we recall (\ref{winding}), we find some other invariant classes
under the winding loops (\ref{Kempe}). We identify a conserved
quantity which is the parity of the total chirality (\ref{chirality})
and briefly discuss some other sectors (\ref{symm} and
\ref{special}). We then enumerate separatly odd and even states by
appropriate transfer matrices, which allow to extract entropies of
infinitely-long strips (section \ref{oddeventm}).  We introduce a
Monte-Carlo dynamics that does not conserve the parity in section
\ref{ergodic}, and compute the fraction of odd states extrapolated to
the thermodynamic limit and the order-parameter in section
\ref{Results}.  In appendix \ref{appA}, in order to check the construction of section \ref{enumeration}, we recall the method of
Pfaffians to enumerate dimer coverings, and give the actual numbers of
configurations. It also allows to
discuss a general invariant of dimer coverings (appendix
\ref{invariantdimer}), which in turn is useful to define another
formulation of the parity (appendix \ref{alternatives}). The appendix
\ref{SM} treats the same problem where the toroidal boundary condition
is replaced by a Klein bottle geometry, leading to similar conclusions.

\section{Model}
\label{Model}
We consider the model of color variables $\sigma_{i}=A,B,C$ defined on
the edges $i=1,\dots,N$ of a regular honeycomb lattice (sites of the
kagome lattice) with periodic boundary conditions. The three edges
meeting at each vertex must be in different colors: this local
constraint defines the 3-color Baxter's model.\cite{Baxter} Each state
is referred as a ``3-coloring'' of the lattice. The number of valid
configurations, \textit{i.e.} respecting the constraints everywhere scales as
$\exp(N s_{\infty})$ in the thermodynamic limit ($s_{\infty}=0.126375...$).\cite{Baxter}

\begin{figure}[h]
\psfig{file=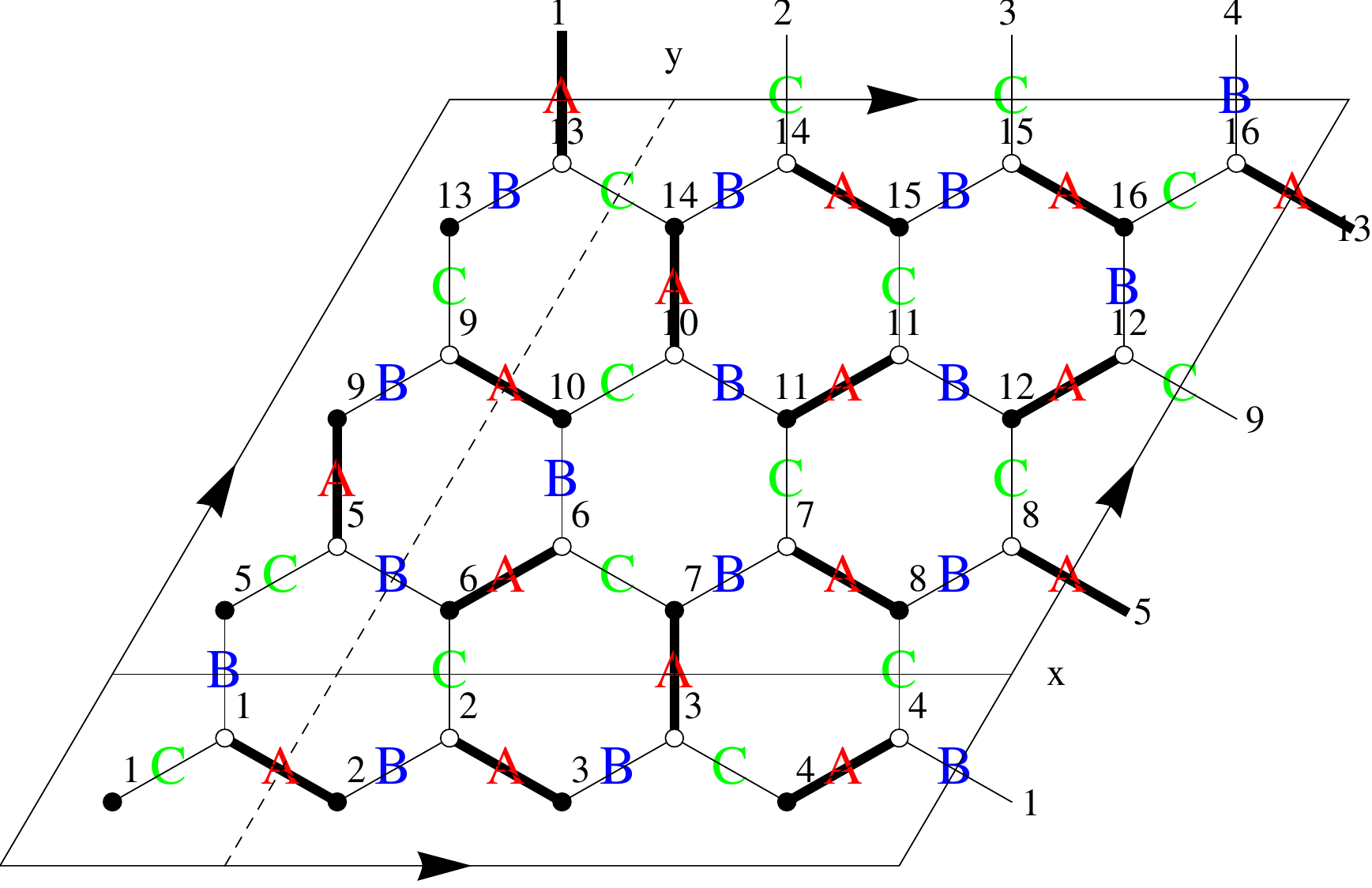,width=8.5cm,angle=-0}
\hspace{1cm}
\caption{Hexagonal lattice with rhombus shape (R) and periodic boundary conditions of the torus geometry. A dimer and color configurations are shown. Here $L=M=4$ and $N=3\times L \times M=48$ edges.  The honeycomb lattice is bipartite and is divided in $L \times M$ black and $L \times M$ white sites. The solid and dashed line show two distinct nonlocal cuts that define topological conserved numbers.}
\label{Honlatt}
\end{figure}

\begin{figure}[h]
\psfig{file=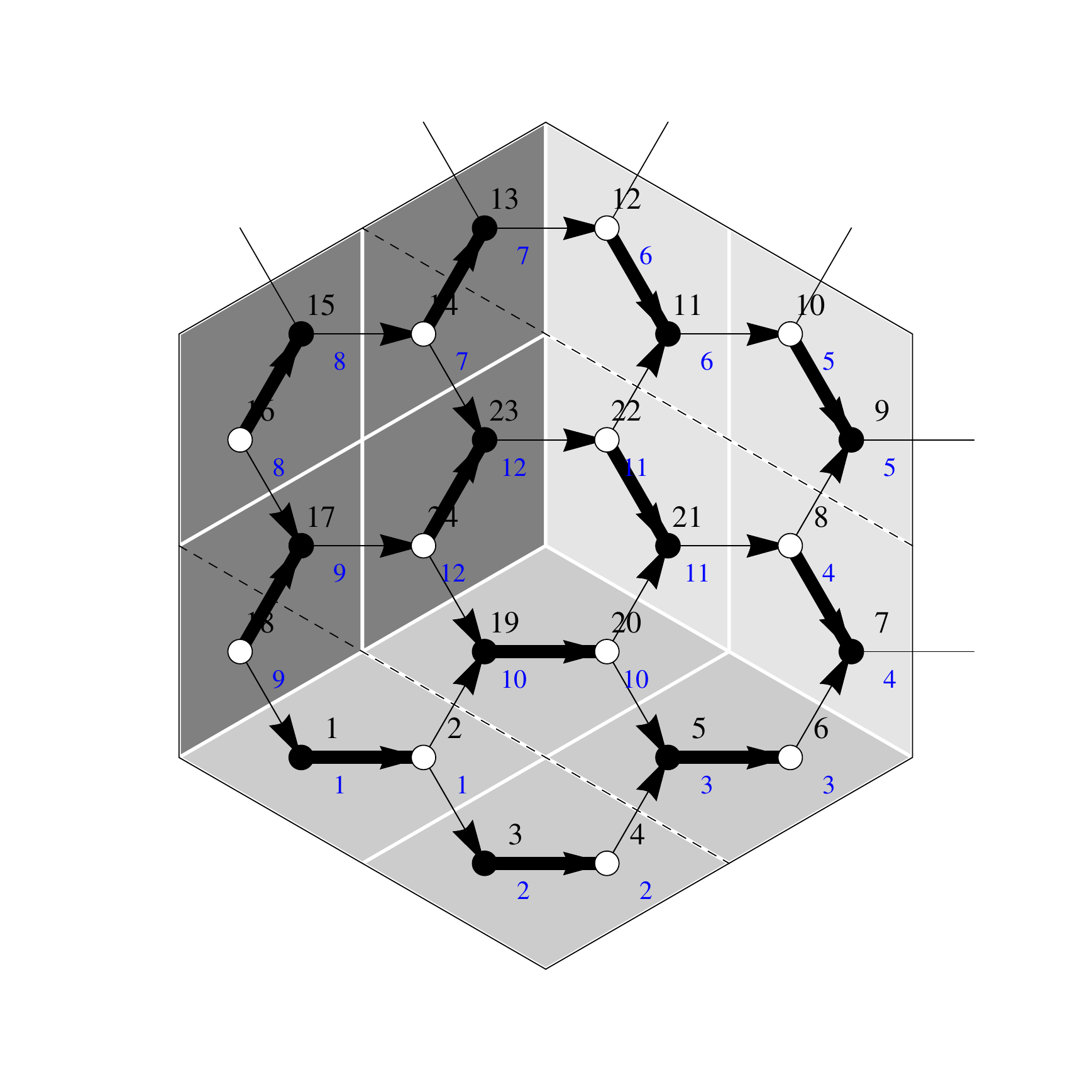,width=8.0cm,angle=-0}
\caption{Hexagonal lattice with hexagonal shape (H), here $L=2$ and $N=9L^2=36$ edges ($N_v=6L^2$). A special dimer configuration is shown, sometimes called the ``empty room'' state in the context of tilings an hexagon with lozenges: the gray colors of the lozenges correspond to the dimer orientations and can be seen as the three walls of an $L^3$ ``empty room''. The dashed line shows one nonlocal cut accross the boundaries.}
\label{Hexlatt}
\end{figure}

In the following, the model is defined on finite-size clusters with
two different shapes, the rhombus (R) and hexagonal (H) shapes shown in
Figs.~\ref{Honlatt} and \ref{Hexlatt}.  We also consider two different
boundary conditions, realizing the topology of the ``torus'' (see the arrows in Fig.~\ref{Honlatt}) or of the
``Klein bottle'' (appendix \ref{SM}).  Both surfaces have zero Euler characteristic and
regular hexagonal lattices fit without introducing noncubic
vertices. 


\section{Exact construction and enumeration of 3-colorings on small lattices}

\subsection{Exhaustive construction}
\label{enumeration}

The exhaustive construction of states satisfying the constraints
everywhere is possible numerically only on very small clusters, since
the number of states increases exponentially with the system size.
For this, we first construct the dimer coverings of the lattice by filling an
empty lattice with dimers, and checking the constraints at each
step.\cite{methodenum} Once a dimer covering is obtained, each vertex
has one edge (out of three) occupied by a dimer -this edge is called
color A. The other two edges form a closed loop of even length (on the
clusters with periodic boundary conditions we have considered), which
is filled with B and C (or C and B) alternatively (see
Fig.\ref{Honlatt} for an example). The dimer configuration has $n_D$
such nonintersecting closed loops, which define $2^{n_D}$
colorings. Note that we obtain all 3-colorings in this way since any
of them can explicitly be decomposed in a dimer configuration (the A
colors) plus a loop configuration. It is numerically more efficient
than enforcing the color constraint on every vertex. By constructing
all dimer coverings, we compute the partition function,
\begin{equation}
Z = \sum_D 1,
\end{equation}
and the number of three-colorings,
\begin{equation}
Z_3 = \sum_D 2^{n_D},
\label{z3def}
\end{equation}
where $n_D$ is the number of loops of the dimer configuration $D$. In
this form, $Z_3$ is also the partition function of the O(2) fully
packed loop model on the honeycomb lattice.  The table \ref{table1} gives the numbers $Z$ and $Z_3$ for the torus 
 boundary conditions, obtained by the
explicit construction of individual states as explained above. $Z_3$
is a multiple of six, since the six color permutations of the
first edge are equivalent, while $Z$ is a multiple of three only for the
torus, where the three edge directions are equivalent.
As a first check, the
entropy per site for $N=192$, $\frac{1}{N} \log Z_3=0.1326$, differs
from Baxter's thermodynamic limit, $0.126375$,\cite{Baxter} by a
typical $1/N$ correction, as expected.  In order to check these
numbers more carefully, we have calculated $Z$ exactly on finite-size
systems by using the method of Pfaffians (details are given in appendix \ref{appA}).  While no closed form is known for
$Z_3$ on finite-size systems and the method of Pfaffians is not applicable, we have
used a numerical transfer matrix method that we explain now.

\begin{table}
\begin{center}
\begin{tabular}{rrrrr}
  \hline \hline
  $N$ & L & T & $Z$ & $Z_3$    \\
  \hline
  9 & 1 & H & 6 & 12  \\
  36 & 2 &  H & 120 & 504   \\
  81 & 3 & H & 15,162 & 135,552    \\
  144 & 4 & H & 13,219,200 & 358,453,104   \\
\hline
  12 & 2 & R & 9 & 24   \\
  27 & 3 & R &  42 & 120     \\
  48 & 4 & R & 417 & 2,160    \\
  75 & 5 & R & 7,623 & 49,416   \\
  108 & 6 & R & 263,640 & 3,226,032   \\
147 & 7 & R & 17,886,144 & 475,299,936    \\ 
192 & 8 & R & 2,249,215,617 & 113,902,581,984  \\ 
  \hline \hline
\end{tabular}
\end{center}
\caption{Number of dimer coverings, $Z$, and 3-colorings, $Z_3$, for system size $N$ (linear size $L=M$) and periodic boundary conditions of the ``torus'' geometry. R/H stands for rhombus/hexagonal shapes shown in Figs.~\ref{Honlatt} and \ref{Hexlatt}. $Z$ is also calculated from Pfaffians in appendix \ref{hexper} and \ref{Rhombtorus}, $Z_3$ from transfer matrix in table \ref{tabletm1} for the rhombus geometry.}
\label{table1}
\end{table}

\subsection{Enumeration of $Z_3$ by transfer matrix}
\label{Tmc}

We have used the transfer matrix method to compute the number of
3-colorings on finite-size systems and check the numbers given above. For
this, we define the state of a horizontal row of $L$ vertical edges (see Fig.~\ref{Honlatt}),
by $\sigma=(\sigma_1,\cdots,\sigma_{L})$ where $\sigma_i=A,B,C$. We
consider two successive rows in the $y$ direction, that we denote by
$\sigma$ and $\sigma^{\prime}$ and define a $3^{L} \times 3^{L}$
transfer matrix $T$ by
\begin{equation}
T_{\sigma \sigma'} = \sum_{\alpha | \sigma \sigma'} 1
\end{equation}
where the sum is over all possible configurations $\alpha$ of the intermediate
set of zig-zag edges compatible with the lower and upper rows $\sigma$ and $\sigma^{\prime}$. 

The
partition function reads
\begin{equation}
Z_3 = \sum_{\sigma_1} \dots \sum_{\sigma_{M}} T_{\sigma_1 \sigma_2} T_{\sigma_2 \sigma_3} \cdots T_{\sigma_{M} \sigma_1}  
=
\mbox{Tr} [ T^{M} ]
\end{equation}
where $M$ is the vertical linear size ($N=3 \times L \times M$), and periodic
boundary conditions of the torus geometry have been used. 
Numerically, to obtain the exact integer number of configurations without rounding errors, we
perform the $M$ multiplications of matrices and compute the trace.\cite{multvsdiag}  Alternatively, we also diagonalize the transfer matrix and we have, 
\begin{equation}
Z_3 = \mbox{Tr} [ T^{M} ]=\sum_{i=1}^{3^L} \Lambda_i^{M}
\label{z3trlambda}
\end{equation}
where $\Lambda_{i=1,\dots,3^L}$ are the (complex) eigenvalues of the
transfer matrix. The diagonalisation method is faster but we have to round $Z_3$ to the nearest integer, which works up to 
$10^{16}$ configurations, \textit{i.e.} one over the machine
precision. This is less than working with integers but we may compute
real quantities for much larger system sizes in the second case.

The total number of edges $N_i^x$ of color $i=A,B,C$ in the row
$\sigma$ is the same in the row $\sigma^{\prime}$; this gives two
independent charges $N_A^x$ and $N_B^x$ (see also section
\ref{dyn}).\cite{Baxter} The transfer matrix then factorizes in
smaller sectors with dimensions $C_{L}^{N_A^x}
C_{L-N_A^x}^{N_B^x}$.  For the largest sector with
$N_i^x=(\frac{L}{3},\frac{L}{3},\frac{L}{3})$ ($L$ a multiple
of three), the dimension typically scales as
$L!/(\frac{L}{3}!)^3 \sim 3^{L}/L$. For $L=12$, the
largest dimension is $C_{12}^4C_8^4=34650$, so that we can compute
all eigenvalues. We also use permutation symmetries to avoid computing
symmetry related sectors by restricting to $N_A^x \leq N_B^x \leq
N_C^x$ and applying a multiplicity factor.

The numbers of three-colorings $Z_3$ are given for different $L, M$ in table~\ref{tabletm1} and match those obtained by the exhaustive construction, given in table~\ref{table1}.
\begin{table}
\begin{center}
\begin{tabular}{rrrrr}
  \hline \hline
  $L$ & $M$  & $Z_3$ & $Z_3^-$    \\
\hline
2 &           2  & 24 & 0 \\
3 &           2  & 48 & 0 \\
3 &           3  & 120 & 60 \\
4 &           2  & 96 & 0 \\
4 &           3  & 408 & 0 \\
4 &           4  & 2160 & 1920 \\
5 &           2  & 192 & 0 \\
5 &           3  & 1284 & 294 \\
5 &           4  & 8208 & 960 \\
5 &           5  & 49416 & 14076 \\
6 &           2  & 384 & 0 \\
6 &           3  & 4752 & 0 \\
6 &           4  & 36096 & 17280 \\
6 &           5  & 317352 & 78120 \\
6 &           6  & 3226032 & 346176 \\
7 &           2  & 768 & 0 \\
7 &           3  & 17412 & 1158 \\
7 &           4  & 185184 & 143808 \\
7 &           5  & 1946964 & 583014 \\
7 &           6  & 30749232 & 4890312 \\
7 &           7  & 475299936 & 424616016 \\
8 &           2  & 1536 & 0 \\
8 &           3  & 68088 & 0 \\
8 &           4  & 916032 & 139776 \\
8 &           5  & 12153168 & 5007360 \\
8 &           6  & 317511600 & 45634752 \\
8 &           7  & 6258486288 & 1333287648 \\
8 &           8  & 113902581984 & 54363353088 \\
9 &           2  & 3072 & 0 \\
9 &           3  & 266232 & 4596 \\
9 &           4  & 4285632 & 3082752 \\
9 &           5  & 80964996 & 31696566 \\
9 &           6  & 3384078480 & 176330736 \\
9 &           7  & 87113393160 & 33997363116 \\
9 &           8  & 2513986458816 & 719824701888 \\
9 &           9  & 84049269591720 & 5365286483676 \\
10 &           2  & 6144 & 0 \\
10 &           3  & 1058808 & 0 \\
10 &           4  & 20484096 & 10229760 \\
10 &           5  & 529208112 & 198732000 \\
10 &           6  & 35145601224 & 2194614720 \\
10 &           7  & 1338325873128 & 1081718221080 \\
10 &           8  & 50904839729376 & 14465622318720 \\
10 &           9  & 2411622439855752 & 463649604519600 \\
10 &          10  & 111152775037945584 & 95192069243340960 \\
  \hline \hline
\end{tabular}
\end{center}
\caption{Number of 3-colorings, $Z_3$, computed by numerical transfer matrix, for rhombi of size $N=3 \times L \times M$ and periodic boundary conditions (Fig.~\ref{Honlatt}). $Z_3^-$ is the number of odd colorings, from section \ref{oddeventm}.}
\label{tabletm1}
\end{table}

\section{Dynamics and conservation laws}
\label{dyn}

The simplest collective dynamics consists of exchanging colors along
loops of sites of two colors (they are closed loops when periodic
boundary conditions are employed). It is the simplest way to preserve
the constraints (note that a single spin flip would not).  This
collective dynamics is also that defined in Monte-Carlo
algorithms\cite{Huse,Moore,Chandra,Chakraborty,CastelnovoMC,Chern} and quantum
three-coloring models.\cite{Castelnovo,Cepas} It is known to be
nonergodic even when all loops are flipped.\cite{Huse,Mohar,Moore} A
different point is that it is also strongly nonergodic (with an
exponential number of sectors) when only the small loops are
flipped.\cite{Cepasq}

Since we have constructed all possible states, we now study why the
dynamics of all loops is nonergodic.  We probe whether two states are
connected by the loop dynamics or not.  For this, we start from a single state in the
ensemble constructed in section \ref{enumeration} and flip all of its $n_l$ loops
(winding and nonwinding), giving $n_l$ possible new states. We iterate
this procedure until no new state is created (see
[\onlinecite{technical}] for technical details). If some states in the
ensemble have not been reached, we then take a new state and reiterate
the same process. When no new state is available in the ensemble, we
are sure that we have constructed all classes of states, closed under
this dynamics.

\subsection{Winding-number sectors}
\label{winding}
The dynamics of \textit{nonwinding} loops conserves some
topological numbers. In the present model, they are obtained by
defining three cuts oriented at 120 degrees, that cut the edges at 90
degrees and go through the centers of the hexagons (see the first two
$x$ and $y$ cuts in Fig.~\ref{Honlatt}). Counting the number of colors
along those cuts is a conserved quantity since any nonwinding loop
intersects twice the cuts in locations where the colors are different.\cite{Castelnovo} Flipping a loop exchanges the
two colors: this gives nine conserved numbers $N_i^{\alpha}=0,\dots,n^{\alpha}$,
$i=A,B,C$, $\alpha=x,y,z$ with some constraints.  We have indeed
\begin{equation}
N_A^{\alpha}+N_B^{\alpha}+N_C^{\alpha}=n^{\alpha}
\end{equation} 
where $n^{\alpha}$ is the number of sites along the cut
 $\alpha$. For example, when $L=M$, we have $n^{\alpha}=L$ for the rhombus shape (Fig.~\ref{Honlatt}), $n^{\alpha}=3L$ for the
hexagonal shape with torus geometry (Fig.~\ref{Hexlatt}). 
So that for each $\alpha$, only two out of three are
independent.  The second constraint comes from the conservation of
these numbers from row to row. Consider the square form $L=M$ for instance. We have,
\begin{equation} 
 L (
N_i^{x}+N_i^{y}+N_i^{z})=\frac{N}{3},
\end{equation}
because the sum over $x,y,z$ can be seen as a sum over the three
sublattices and $N/3$ is the total number of sites with color
$i$. This leads to $\sum_{\alpha} N_i^{\alpha}=L$ or $3L$, depending on the shape. This
constraints the third direction to be determined from the first two.
Therefore, only four of the nine $N_i^{\alpha}$ are independent.  But
to emphasize the symmetries it is convenient to write a general
configuration of a topological sector by
$N_i^{\alpha}=(a_x,b_x,c_x,a_y,b_y,c_y,a_z,b_z,c_z)$. Since the number of
colors is always positive, we also have inequalities such as
$N_A^{\alpha}+N_B^{\alpha} \leq n$ (for $L=M$) which further reduces the number of
possibilities to $(\sum_{k=1}^{n+1} k)^2=[(n+1)(n+2)/2]^2$. Some of them are not allowed
however, so the number of sectors is strictly less that that.

\subsection{Kempe sectors and odd/even classes}
\label{Kempe}

When \textit{winding} loops are added in the dynamics, the winding
numbers are no longer conserved and topological sectors are \textit{a
  priori} connected. In fact, we find some disconnected classes,
sometimes called ``Kempe'' classes in the
literature.\cite{Mohar,Belcastro} The number of classes we find is
given by $n_K$ in table \ref{tableKT}, up to some degeneracies that we
indicate and discuss in section \ref{symm}. $n_K$ depends on the
geometry and it increases with the system size. Given the limitation
in sizes, it is an open question as to whether it is infinite in the
thermodynamic limit.

\begin{table}
\begin{center}
\begin{tabular}{rrp{7cm}}
  \hline \hline
  $N$  & $n_K$  & $Z_{i=1,\cdots,n_K} (\pm)$  \\
  \hline
  9 &  2 & 6(+), 6(-) \\
  36 & 2 &  360(+), 144(-) \\
  81 & 3 & 111,840(+), 60(-), 23,652(-) \\
  144 & 5 & 315,051,888(+), 42,953,088(-), 44,928(x6)(-), 155,520(-), 23,040(-) \\
\hline
  12 & 1 & 24 (+)\\
  27 & 2 & 60(+), 60(-)  \\
  48 & 2 & 240(+), 1,920(-) \\
  75  & 3 & 35,340(+), 276(-), 13,800(-)  \\
  108 & 3 & 2,879,856(+), 307,296(-), 19,440(x2)(-) \\
147 &  4 & 50,683,920(+), 1140(-), 424,598,328(-), 6 (x2758)(-)  \\ 
192 &  4 & 59,539,228,896(+), 54,178,583,040(-), 14,555,136(-), 28,369,152(x6)(-)  \\ 
  \hline \hline
\end{tabular}
\end{center}
\caption{Number of invariant classes $n_K$ and number of states in each class $Z_{i=1,\dots,n_K}$ ($\sum_{i=1}^{n_K} n_i Z_i=Z_3$, where $n_i$ is the sector multiplicity, given in brackets). The even/odd chirality of each class is given by $\pm$.}
\label{tableKT}
\end{table}

Each Kempe sector contains several winding sectors connected by
the motion of the winding loops. In the torus geometry, distinct Kempe
sectors (we call ``distinct'' two sectors that are not related by the loop
motion but also by a lattice symmetry) contain distinct winding
sectors, but this is not true in the Klein bottle geometry.\cite{192} 

The number of states in each sector is given by $Z_i$ with
$i=1,\dots,n_K$ and varies from $Z_i \lesssim Z_3$ down to $Z_i=6$ states. This
reflects the topological sectors themselves, which are expected to
have Gaussian distribution.\cite{Chakraborty}

We now show that we can distinguish some of these sectors by the parity of the chirality and some lattice symmetries.

\subsubsection{Conservation law : odd/even chirality}
\label{chirality}

\begin{figure}[h]
\psfig{file=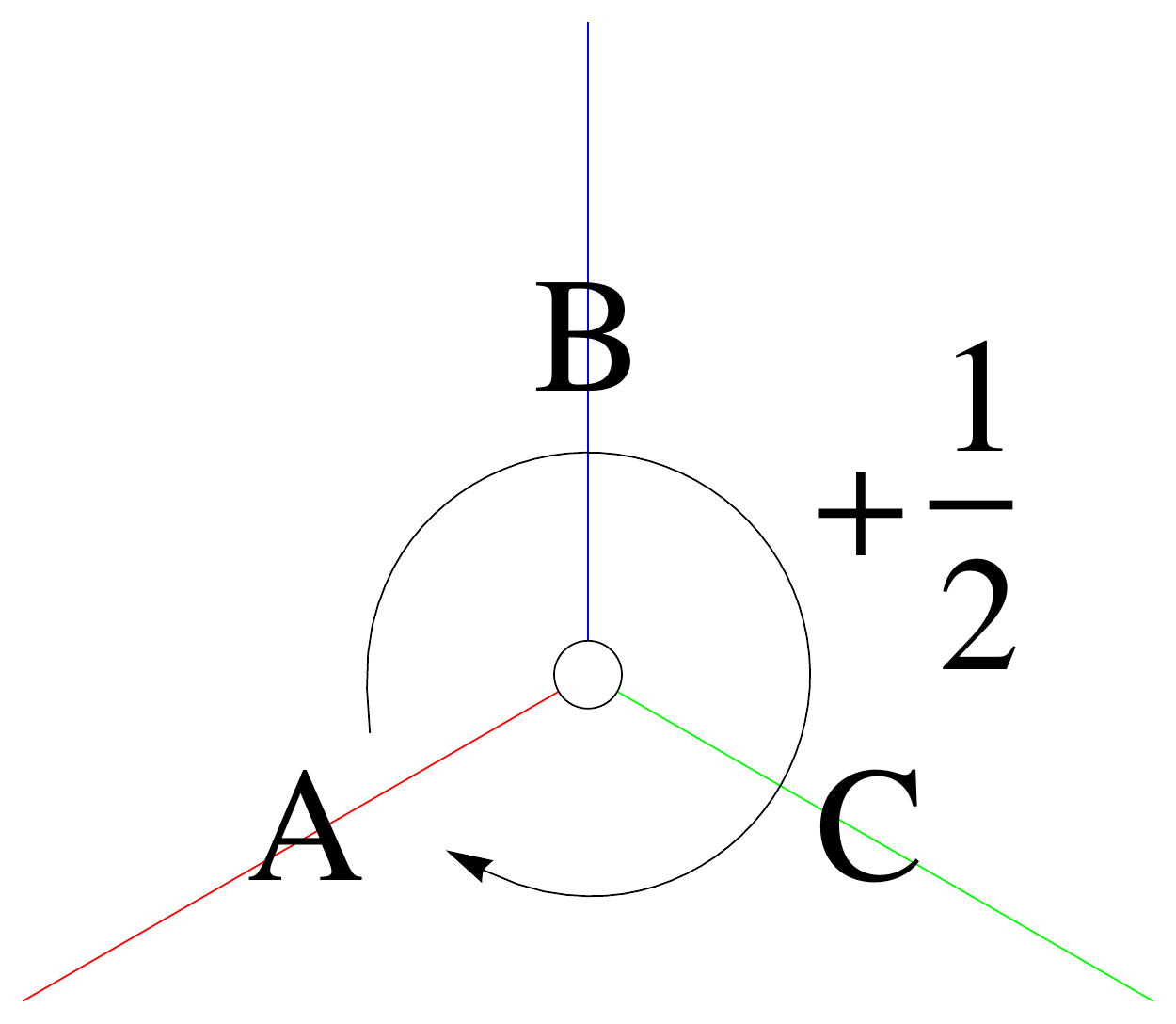,width=3.0cm,angle=-0} \hspace{1cm}
\psfig{file=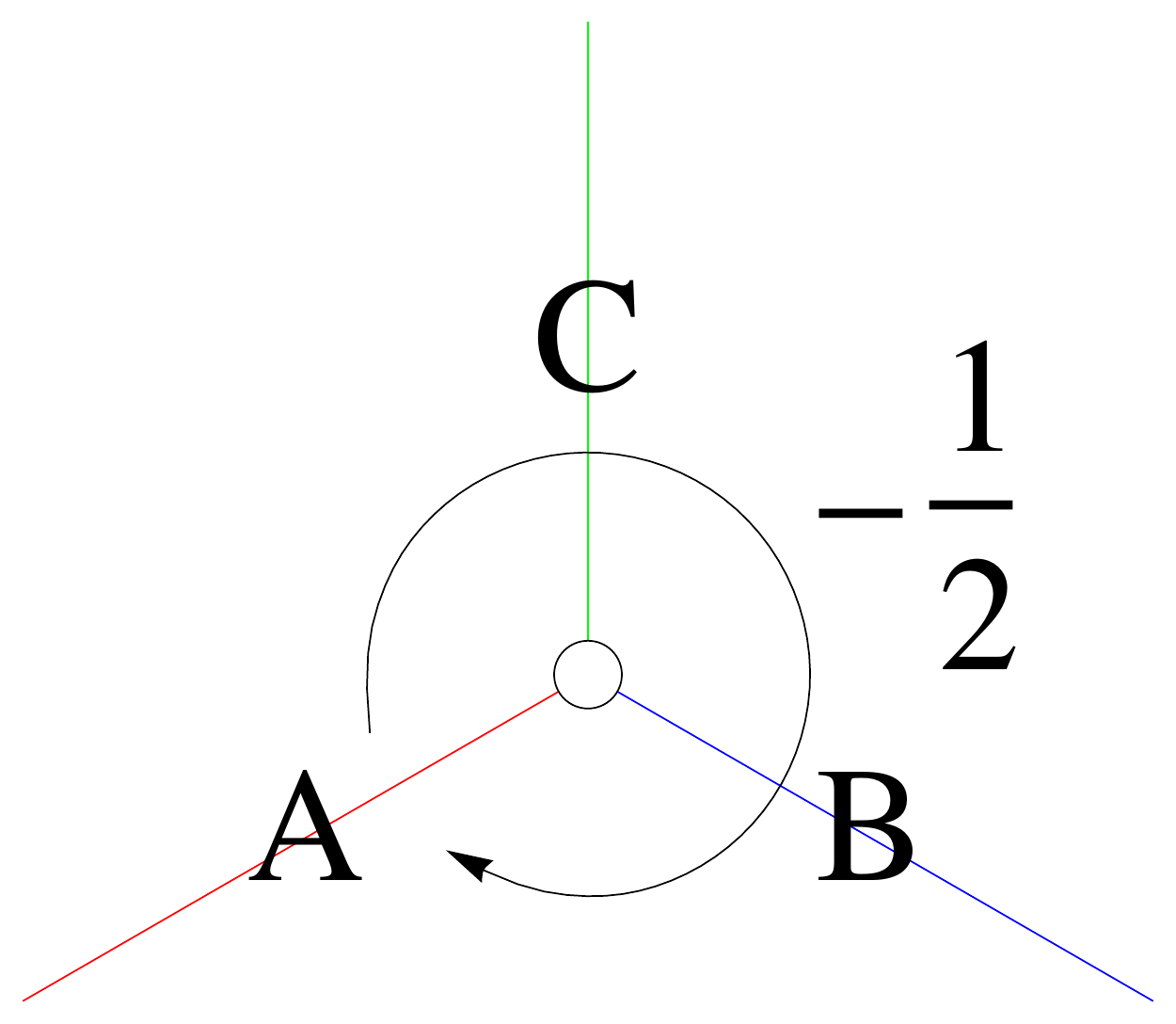,width=3.0cm,angle=-0} \hspace{3cm} 
\caption{Definition of the chirality at a given (white) vertex (the definition is the same for black vertices). The sum of the chirality of all vertices (black and white) can be even or odd and its parity is conserved.}
\label{chiralitydef}
\end{figure}

We define the total chirality of a state by,
\begin{equation}
m= \sum_v \chi_v,
\end{equation}
where $\chi_v=\pm 1/2$ for an ABC (or ACB) orientation of the edges by
turning clockwise around any vertex $v$ (Fig.~\ref{chiralitydef}). 
The sum is over all (black and white) $2N/3$ vertices of the bipartite honeycomb lattice.
Since the total number of vertices is even, $m$ is an integer that can be even or odd.

First, the system always has the chiral symmetry between $m$ and $-m$, obtained by exchanging all A and B, which can be done by moving all A-B loops.
Second, nonwinding loops have zero chirality so that $m$ is conserved under the dynamics of \textit{nonwinding} loops.
Third, there may be a finite chirality along a winding loop,
$m_l$ where $l$ is the length of the loop, which is always even. Flipping this loop reverses the chirality of all its vertices, so $m_l \rightarrow -m_l$. Since $l$ is even, $m_l$ is an integer, and the total chirality changes by $2m_l$ which is an even integer, so that the parity of $m$,
\begin{equation}
I_3=(-1)^{m},
\end{equation}
 is conserved in the dynamics of all loops. This defines two odd/even classes and we have labelled the Kempe sectors by this parity in table \ref{tableKT}. Note that the odd class is further split into sectors, some of which we will discuss below. Note also that on clusters with open
 boundary conditions (cylinder or plane), there are open loops at the boundaries
 that change the chirality of an odd number of vertices, and thus do
 not conserve the parity.

We give two concrete examples of states with odd or even chirality. We consider
first a periodic state with a tripled unit-cell (called $\sqrt{3}
\times \sqrt{3}$), obtained by stacking A-B-C along any of the three
lattice directions. While it is always compatible with the hexagonal
shape, $L$ and $M$ have to be multiples of three on the rhombus shape
and it does not fit in the Klein bottle geometry. For this state, the
total chirality is zero, always even. For the ``$Q=0$'' state, all
(black and white) vertices have the same configuration, say ABC, so
that $m=N/3$. For the square shape with $L=M$, $m$ has the parity of
$L$. Therefore for $L$ odd and multiple of three (so as to fit the
$\sqrt{3} \times \sqrt{3}$) we have at least two independent
sectors. The conclusion $n_K>1$ was already reached in
Ref.~\onlinecite{Mohar} and extended for even $L$.

We describe in the appendix \ref{alternatives} some other formulations of the same invariant.

\subsubsection{Lattice symmetries}
\label{symm}
We find some degenerate sectors with an identical number of states (table \ref{tableKT}). This degeneracy can be explained by
lattice symmetries. 

Permutation of colors can be implemented by a loop motion, so that two
topological sectors related by permutation symmetry, \textit{e.g.}
$(a_x,b_x,c_x,a_y,b_y,c_y,a_z,b_z,c_z)$ and
$(b_x,a_x,c_x,b_y,a_y,c_y,b_z,a_z,c_z)$ belong to the same Kempe
sector.  Permutations generate at most six sectors.

Applying lattice symmetries is a different
operation that generates other sectors that may not be connected by
the dynamics. For instance, applying a $x-y$ mirror plane
gives $(a_y,b_y,c_y,a_x,b_x,c_x,a_z,b_z,c_z)$. Successive applications of the
three mirror planes generates at most six sectors if all the charges
are different. 

 For instance, for $N=108$, there are two Kempe sectors with 19440
 states. They are related by a mirror symmetry. Indeed the winding
 sectors in one of these two Kempe classe have the special form
 $(a,b,c,b,c,a,c,a,b)$ (up to color permutations), and
 $(b,c,a,a,b,c,c,a,b)$ in the other. Mirror symmetries up to
 permutations generate only two sectors in this case. However, for
 $N=144$ or $N=192$, there is a six-fold degeneracy that is obtained
 by using the three mirror planes.

 These degeneracies are robust in that any local
 perturbation that breaks the mirror symmetry has the same average in
 both sectors, as expected for topological sectors.\cite{Ioffe}

\subsubsection{Special sectors}
\label{special}

Some sectors have no weight in the thermodynamic limit.

A first example is the sector of the ``$Q=0$'' state, when $L$ is
odd. This state belongs to the smallest nondegenerate sector (see table
\ref{tableKT}). It has maximal chirality with all vertices in the
$\pm$ state, $m= \pm N/3$.  Since $m$ is a multiple of $L$, and all
winding loop lengths are multiples of $2L$, the chirality changes by
$2L$ and remains therefore a multiple of $L$. The new winding loops
have lengths that remain multiples of $2L$ so that the Kempe sector
has $m=\pm (1+2n)L$ which is not only odd but is also a multiple of
$L$, a property that remains stable in the dynamics.

A second example is found for $N=147$.  We find 2758 degenerate sectors containing only 6
states. Some of these sectors are related by translation symmetry but
are special in that they minimize the number of loops, \textit{i.e.} three. Each
loop, say the A-B one, takes all the $N/3$ A edges and the $N/3$ B edges. This
loop connects all sites of the honeycomb lattice and is an Hamiltonian
cycle. Flipping an A-B loop exchanges the A-C and B-C loops and
therefore keeps the lengths identical. Such a structure is therefore
 stable and has only six states obtained by the only six possible permutations. We have not found this
structure on the other clusters available, but it may exist for larger sizes.  Note
that while there is always a Hamiltonian path on a regular honeycomb
lattice, this is not always true for such three intertwined
Hamiltonian paths.

\section{Enumeration of odd/even states by transfer matrix}
\label{oddeventm}
We compute the number of odd/even states by introducing a modified partition function, $P(z)$ with a fugacity $z$,
\begin{equation}
P(z) = \sum_C z^{m} 
\label{partitionmod}
\end{equation}
where the sum is over all 3-color configurations and $m$ is the total chirality of the configuration $C$, defined in section \ref{chirality}. The total number of colorings is 
\begin{equation}
Z_3=P(1).
\end{equation} 
However, $z$ in Eq.~\ref{partitionmod} can be any number. Given that, by symmetry, there is the same number of states with $m$ and $-m$, we have $P(1/z)=P(z)$. In particular, we consider $P(z)$ with $z=-1$, which contains terms of the form $(-1)^m=\pm 1$. These terms differ by their signs in the two parity sectors (\ref{chirality}), therefore all even colorings are counted with a $+$ sign while odd colorings are counted with a $-$ sign. We obtain the number of odd colorings by 
\begin{equation}
Z_3^- = \frac{1}{2}[P(1)-P(-1)]. 
\label{z3mi}
\end{equation}
and the number of even colorings, from $Z_3=Z_3^++Z_3^-$. To compute this number, we construct a transfer matrix with elements,
\begin{equation}
T_{\sigma \sigma^{'}}(z)=\sum_{\alpha | \sigma \sigma^{'}} z^m,
\end{equation} 
 where the sum is over the configurations $ \alpha$ of the
 intermediate zig-zag row compatible with the pair of rows $\sigma$
 and $ \sigma'$, and $m$ is the partial chirality of that row. We
 restrict the fugacity to $z=-1$, so that the transfer matrix has
 integer entries.  The number of configurations $Z_3^-$ is
 then computed as an integer by matrix multiplications, from 
\begin{equation}
P(\pm  1)=\mbox{Tr}[T(\pm 1)^{M}],
\end{equation} 
and   Eq.~\ref{z3mi}.
 The results are given
 in table~\ref{tabletm1} and they match the
 numbers found by the dynamical process (table \ref{tableKT}). 

\subsection{Rotation symmetry}

We note that the transfer matrix $T(-1)$ has an
exact SU(3) symmetry even on finite size systems, since we have explicitly checked that
\begin{equation}
[T(-1), \sum_{i=1}^{L} \lambda_c(i)]=0
\end{equation}
for all the $c=1,\dots,8$ generators of SU(3), $\lambda_c(i)$ on edge $i$. A
consequence is that the eigenvalues of $T(-1)$ have degeneracies that are that of the multiplets
of SU(3). In this language, the conserved isospin and hypercharge are obtained from the number of colors, by
 $T_3=\frac{1}{2}(N_A^x-N_B^x)$ and $Y=\frac{1}{3}(N_A^x+N_B^x-2N_C^x)$.

\begin{figure}[h]
\psfig{file=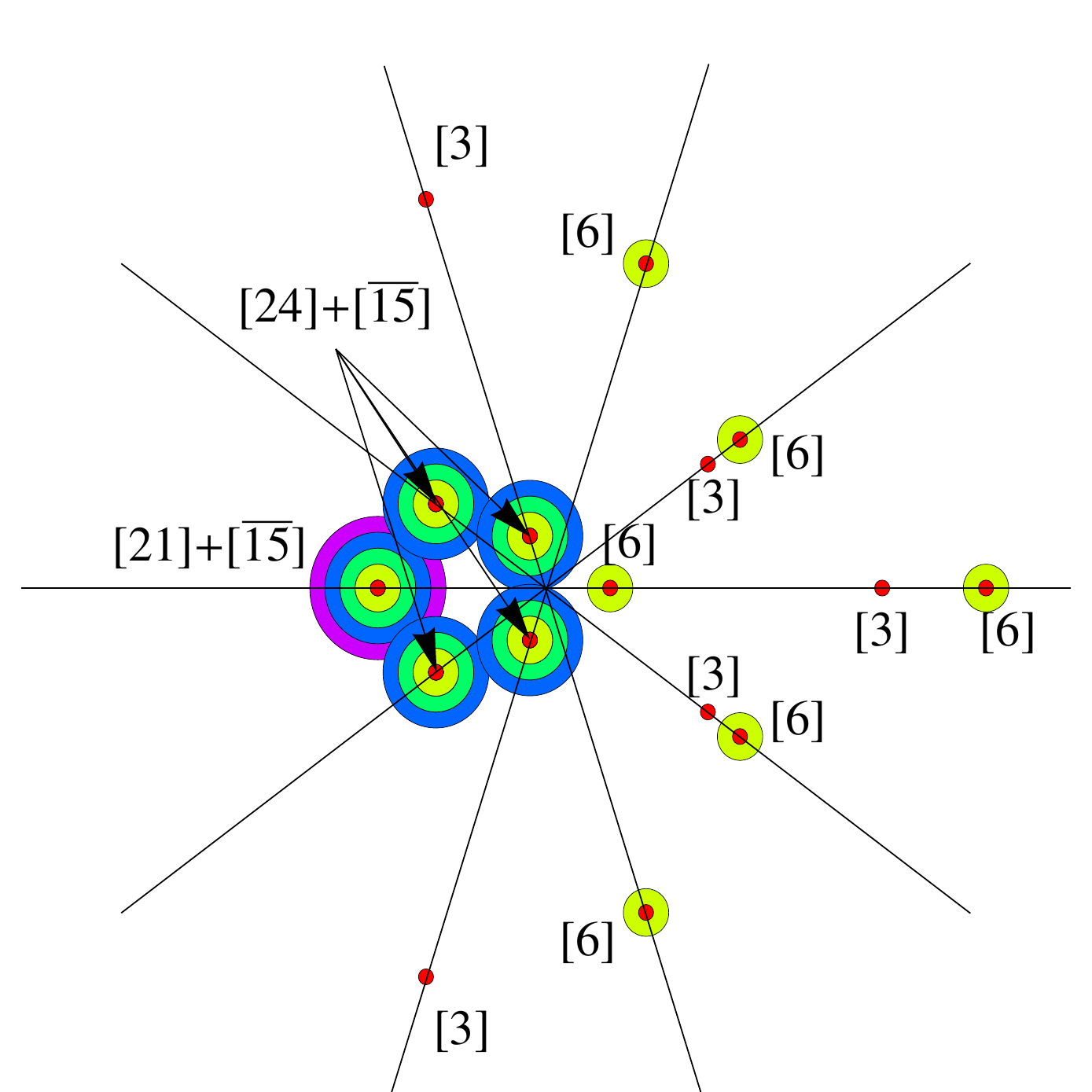,width=7.0cm,angle=-0}
\caption{Eigenvalues of the transfer matrix $T(-1)$ (given here in the complex plane for $L=5$) are degenerate and form the multiplets of SU(3), $[3], [6], [\bar{15}], [21], [24]$. The colors indicate different charge sectors (the degeneracy due to permutation-related sectors is not shown). }
\label{eigenvalues}
\end{figure}

Fig.~\ref{eigenvalues} gives an example of the eigenvalues of $T(-1)$
for $L=5$, in the complex plane. We find that they form multiplets belonging to the following irreducible representations of
SU(3), $ 5 \times [3]+ 6\times [6] + 5\times [\bar{15}] +[21] +4
\times [24] \equiv 3^5 $, where the number in bracket gives the
dimension of the representation, the bar specifies the right of left
representation and the number in front the number of such
multiplets. Some additional degeneracies are found, such as $[24]$ and $[\bar{15}]$, or $[21]$ and $[\bar{15}]$, which seem to be 
accidental but may indicate a higher
symmetry than SU(3).

This symmetry enhancement at a critical point
($z=-1$), is very similar to a spin-ice model where an exact SU(2)
symmetry was found in finite-size systems.\cite{Jaubert} It was also
conjectured that the SU(3) symmetry holds for $z=1$ but only in the
thermodynamic limit.\cite{Read,Kondev}

\subsection{Fraction of odd states}

\begin{figure}[h]
\vspace{-.3cm}
\psfig{file=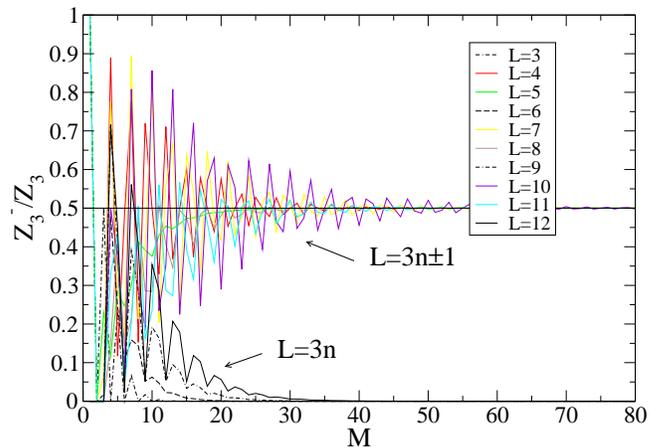,width=9.5cm,angle=-0}
\vspace{-.8cm}
\caption{Fraction of odd states as a function of $M$ for a strip of size $N=3 \times L \times M$ with rhombus shape and periodic boundary conditions, from transfer matrix. When $M \rightarrow \infty$, the fraction goes to 0 for $L=3n$  and 1/2 for $L=3n \pm 1$, thus exhibiting a $Z_2$ symmetry between even and odd states.}
\label{nminustm}
\end{figure}

We compute the exact fraction of odd states for finite-size $L$, as a real number, from the complete diagonalisation of
the transfer matrices,
\begin{equation}
\frac{Z_3^-}{Z_3} = \frac{1}{2}\left[1-\frac{P(-1)}{P(1)} \right] 
\end{equation}
The result is given in Fig.~\ref{nminustm} as a function of $M$, for different $L$ up to $L=12$.

We can study the limit of large transverse size $M
\rightarrow +\infty$, \textit{i.e.} a thermodynamic limit with zero aspect ratio $r=L/M
\rightarrow 0$. In this case, we are interested in the
largest eigenvalues, and we have to
distinguish two cases, depending on whether $L$ is a multiple of three or not.

\subsubsection{$L=3n$ ($L$ a multiple of three)}

\begin{figure}[h]
\vspace{-.3cm}
\psfig{file=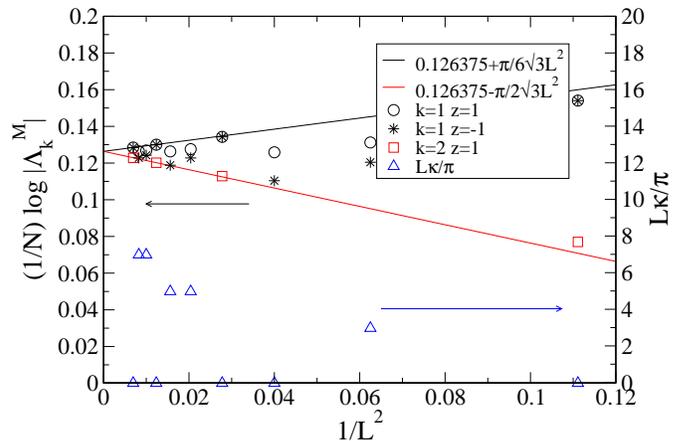,width=9.5cm,angle=-0}
\vspace{-.8cm}
\caption{First eigenvalues of $T(1)$ and $T(-1)$ and the angle $\kappa$ defined by $\Lambda_1^-=|\Lambda_1^-|e^{\pm \imath \kappa}$. The lines are finite-size corrections to Baxter's thermodynamic limit, $s_{\infty}=0.126375$, predicted from conformal invariance.}
\label{eig}
\end{figure}

In this case, we find that the largest eigenvalue of $T(1)$, $\Lambda_1$ equals that of $T(-1)$, $\Lambda_1^-$ for the sizes considered ($L=3,6,9,12$, see Fig.~\ref{eig}).
 The eigenvalue is real and belongs to the largest nondegenerate sector with conserved charges $(\frac{L}{3},\frac{L}{3},\frac{L}{3})$, and the eigenvector of $T(-1)$ is a \textit{singlet} of SU(3). We have
\begin{equation}
P(\pm 1) = \Lambda_1^M + \dots,
\end{equation}
so that the contribution in $Z_3^-$ (Eq.~\ref{z3mi}) from the first eigenvalues exactly cancels and we are left with 
\begin{equation}
Z_3 \sim \Lambda_1^{M}; \hspace{1cm} Z_3^- \sim 3 \Lambda_2^{M}
\label{ent}
\end{equation}
where $\Lambda_2<\Lambda_1$ is the second largest eigenvalue of
$T(1)$ which belongs to a sector that is six times degenerate (hence the factor 3 in front), given in Fig.~\ref{eig}. In this case,  the fraction $Z_3^-/Z_3 \sim (\Lambda_2/\Lambda_1)^M$ goes exponentially to zero, as found
in Fig.~\ref{nminustm}. 

The total entropy per site, and the entropy of the odd states read
\begin{eqnarray}
S &\equiv& \frac{1}{N} \log Z_3 = \frac{1}{N} \log |\Lambda_1^{M}| \\ 
S^- &\equiv& \frac{1}{N} \log Z_3^- = \frac{1}{N} \log |\Lambda_2^{M}| 
\end{eqnarray}
when $L$ is a multiple of three, so that $S^-<S$.
From conformal invariance that is expected from the height mapping,\cite{Huse,Kondev}
one expects finite-size corrections,
\begin{equation}
S= s_{\infty}+ \frac{\pi c}{6 \zeta L^2} +o(L^{-2})
\end{equation}
with conformal charge $c=2$, geometrical factor $\zeta=2\sqrt{3}$, and $s_{\infty}=0.126375$ is Baxter's exact result. Since the system is critical, the gap between the first and the second eigenvalue is expected to close with a universal correction,
\begin{equation}
S^--S = -\frac{\pi}{ \zeta L^2} \eta +o(L^{-2})
\label{entropyodd}
\end{equation}
where $\eta=4/3$ is the scaling dimension of the spin-spin correlation function, which is known\cite{Huse,Kondev,Chakraborty} to decay algebraically as $r^{-\eta}=r^{-4/3}$ (see also Fig.~\ref{orderparameter}). These curves are shown in Fig.~\ref{eig} and fits well the data for $L$ a multiple of three, with $\eta=4/3$. Since the second eigenvalue gives the entropy of the odd states, we thus expect that the two classes have the same entropy in the thermodynamic limit. This will be confirmed independently in section \ref{ergodic}.
It is only at zero aspect ratio and $L$
a finite multiple of three that odd states have a smaller entropy than even states, given by the squares in Fig.~\ref{eig} and well-approximated by Eq.~\ref{entropyodd}.

\subsubsection{$L=3n\pm 1$ ($L$ not a multiple of three)}

The first eigenvalue $\Lambda_1$ belongs to a triplet sector with conserved charges $(\frac{L}{3},\frac{L}{3},\frac{L}{3}) \pm (\frac{2}{3},-\frac{1}{3},-\frac{1}{3})$, so that  $\Lambda_1$ is three times degenerate. $\Lambda_1^-$ belongs to the $[3]$ or $[\bar{3}]$ triplet representation of SU(3) (except for $L=5$).
It is found to be different from $\Lambda_1$ with $|\Lambda_1^-|<\Lambda_1$ (compare the stars with open circles in Fig.~\ref{eig}), so that at first order
\begin{equation}
Z_3 \sim 3 \Lambda_1^{M}; \hspace{1cm} Z_3^- \sim \frac{3}{2}\Lambda_1^{M},
\label{entnot3}
\end{equation}
and $Z_3^-/Z_3$ goes to 1/2 when $M \rightarrow + \infty$ for infinitely long strips (as shown in Fig.~\ref{nminustm}). In this
case not only the entropies are identical but also there is the same
number of even and odd states: a $Z_2$ symmetry occurs.  

Whereas $\Lambda_1$ must remain real (since
$P(1)$ is a positive integer), $\Lambda_1^-=|\Lambda_1^-|e^{\imath \kappa}$ is complex
in general (except for $L=5$). In this case, the complex-conjugate eigenvalue also occurs because
$P(-1)$ is an integer (positive or negative). Thus the leading correction to the fraction of odd states
reads
\begin{equation}
\frac{Z_3^-}{Z_3} \sim \frac{1}{2} - \alpha \left(\frac{|\Lambda_1^-|}{\Lambda_1}\right)^{M} \cos(\kappa M) + \dots
\label{oddf}
\end{equation}
where $\alpha$ is the ratio of degeneracies and
$|\Lambda_1^-|<\Lambda_1$.  The convergence has oscillations
in addition to the exponential decrease (see Fig.~\ref{nminustm}).  It is peculiar that there are special
aspect ratios where $Z_3^-/Z_3$ is very close to $1/2$ even for small
$M$. This is the case of $L=4$ for $r=4/14 (0.4999824)$, $4/18 (0.49999948)$, $4/22 (0.499999984)$ \textit{etc.} (the
larger the denominator, the closer to 1/2) and $L=10$ with $10/25 (0.49938)$,
$10/35 (0.499969)$, $10/45 (0.4999984)$, \textit{etc.} These aspect ratios are of the form $2/p$ where $p$
is an odd integer. They correspond in fact to a cancelation of the oscillating
part of Eq.~\ref{oddf}, $\kappa M =\pi/2$ mod $\pi$. The convergence is then
controlled by the next ratio of eigenvalues and is much faster.

It is a question as to whether in the thermodynamic limit aspect
ratios other than zero may have a perfect Z$_2$ symmetry. Partly to address this
question, we need to have access to larger system sizes and we will
use a Monte-Carlo method with a modified algorithm.

\section{Ergodic Monte-Carlo algorithm}

\begin{figure}[h]
\psfig{file=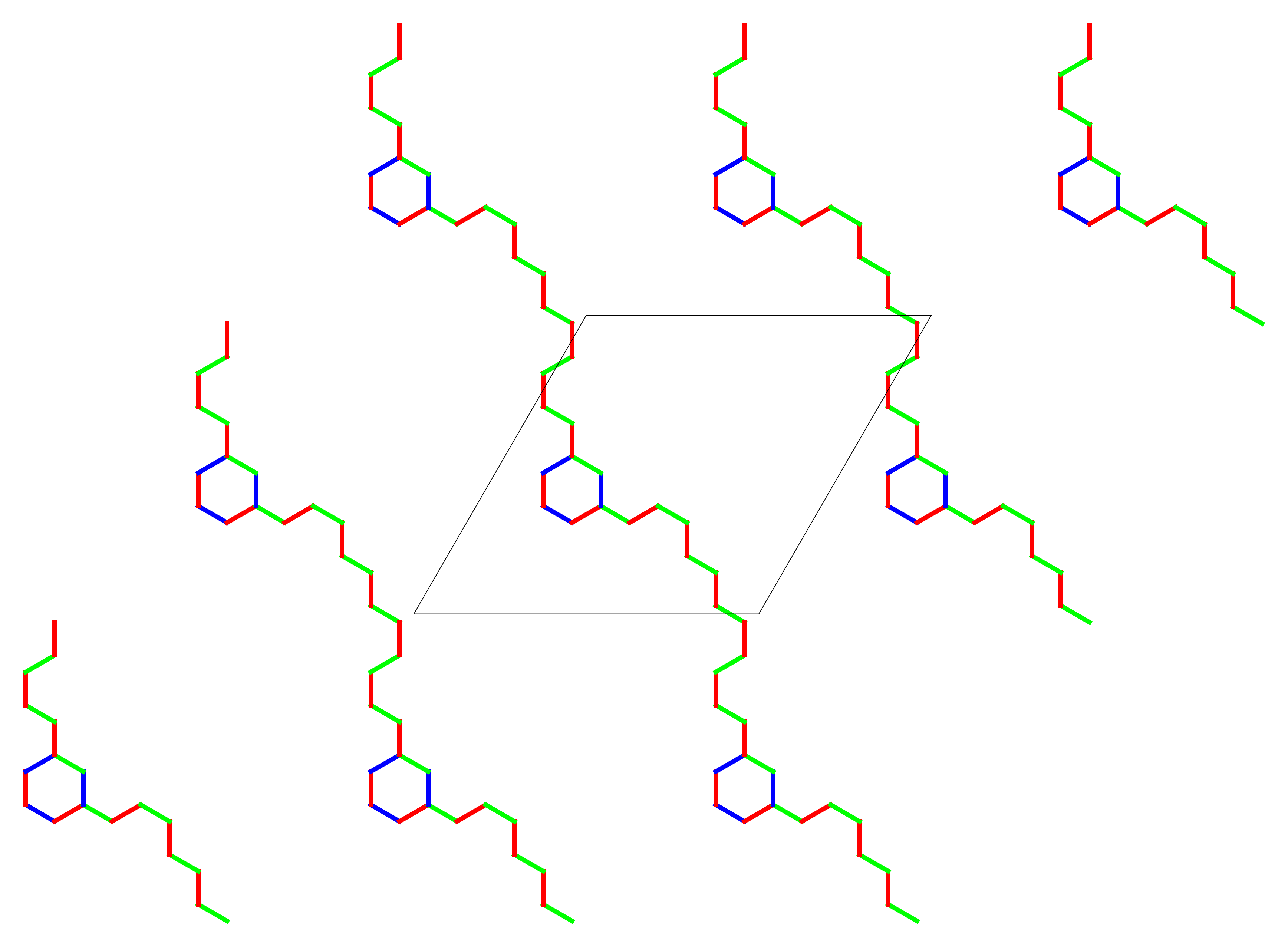,width=7.0cm,angle=-0} \hspace{1cm}
\psfig{file=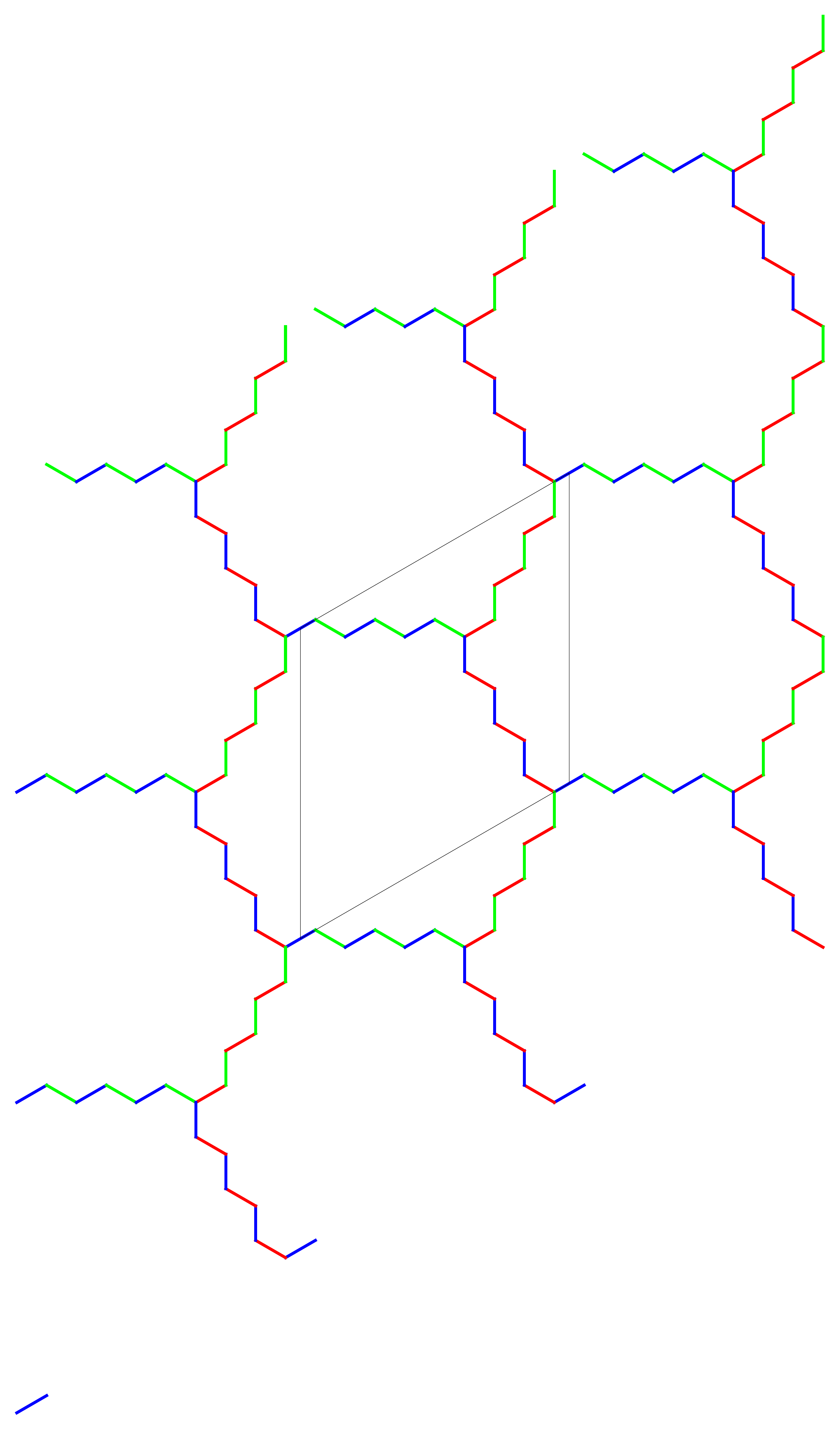,width=5.0cm,angle=-0} 
\caption{Triple-stranded loops make the algorithm ergodic by allowing
  parity changes. Black lozenges are the unit-cell, the loops are
  repeated outside the unit-cell for clarity (top, $N=108$, bottom
  $N=81$).}
\label{overlaps}
\end{figure}

We introduce a Monte-Carlo algorithm that does not conserve the parity
of the chirality, by including the flip of ``stranded'' loops (examples are given in
Fig.~\ref{overlaps}).  We explicitly ckeck that this algorithm can
reach all states on the clusters considered.

\subsection{Algorithm}
\label{ergodic}

In dimer models, loop Monte-Carlo algorithms that include all winding and nonwinding loops are
ergodic. Indeed, given two dimer coverings, we can construct the
``transition graph'', which consists of superimposing the two dimer
coverings. The transition graph gives an ensemble of closed loops and
individual dimers for edges both occupied by a dimer. Flipping all of
its loops takes one state to the other, so that the states are
connected by the motion of the loops. 

In the three color model, the ``transition graph'' argument
does not work. While it is possible to do the same operation for egdes
in the, say, A state, this leaves the right loop configuration for the
B-C sites (as explained in section \ref{enumeration}), with some sites
already occupied and some empty, those that were previously A. The problem
is that the empty sites cannot always be assigned a B or a C, since
they are sometimes surrounded by exactly one B and one C: any
assignement would violate the constraint. A way to solve the problem
is to reorganize segments of the B-C loops, this is possible but it does not
correspond to a loop motion.

In order to give a more clear picture of this, we have looked at the
maximal overlap of two states belonging to two different sectors (this
would give a single loop in the dimer problem). Two representative
examples are given in Fig.~\ref{overlaps}: there we show
only the edges that differ in color between the two states. In both cases, the loop is
``stranded'': it forks at a given vertex into two segments that
recombine at another vertex, either forming a short closed loop (an
hexagon in the top figure) or a winding loop (bottom figure). An
algorithm that would include these special moves will thus be able to
bring the system from one sector to another.

A way to implement numerically these collective flips is through the
introduction and annihilation of defects, \textit{i.e.} local
violations of the constraint.  Recall first that flipping a two-color
loop is equivalent to introducing \textit{two} defects and annihilate
them. First exchange the colors A-B of two neighboring edges. This creates a
pair of defects with two neighbors in the same color, say A-A and
B-B. Propagating them away from each other on the A-B loop by
successively swapping A-B pairs, and recombining them at the end, exchanges the two
colors of the whole loop. There are six such defects\cite{Moore} $q_i$
and $\bar{q}_i$, $i=A,B,C$: they are vertices with, among the three
edges, one edge in color $i$ and two in the same color, thus violating
the local constraint. Since two colors are available, they are denoted
by $q$ or $\bar{q}$. With this notation, an A-B swap creates a pair of
conjugate defects, which can
propagate on the A-B loop and recombine,
\begin{eqnarray}
&& \emptyset \rightarrow q_C+\bar{q}_C \\
&& q_C+\bar{q}_C \rightarrow \emptyset \label{laststep}
\end{eqnarray}
This process is the standard flip of an entire A-B loop.

\begin{figure}[h]
\psfig{file=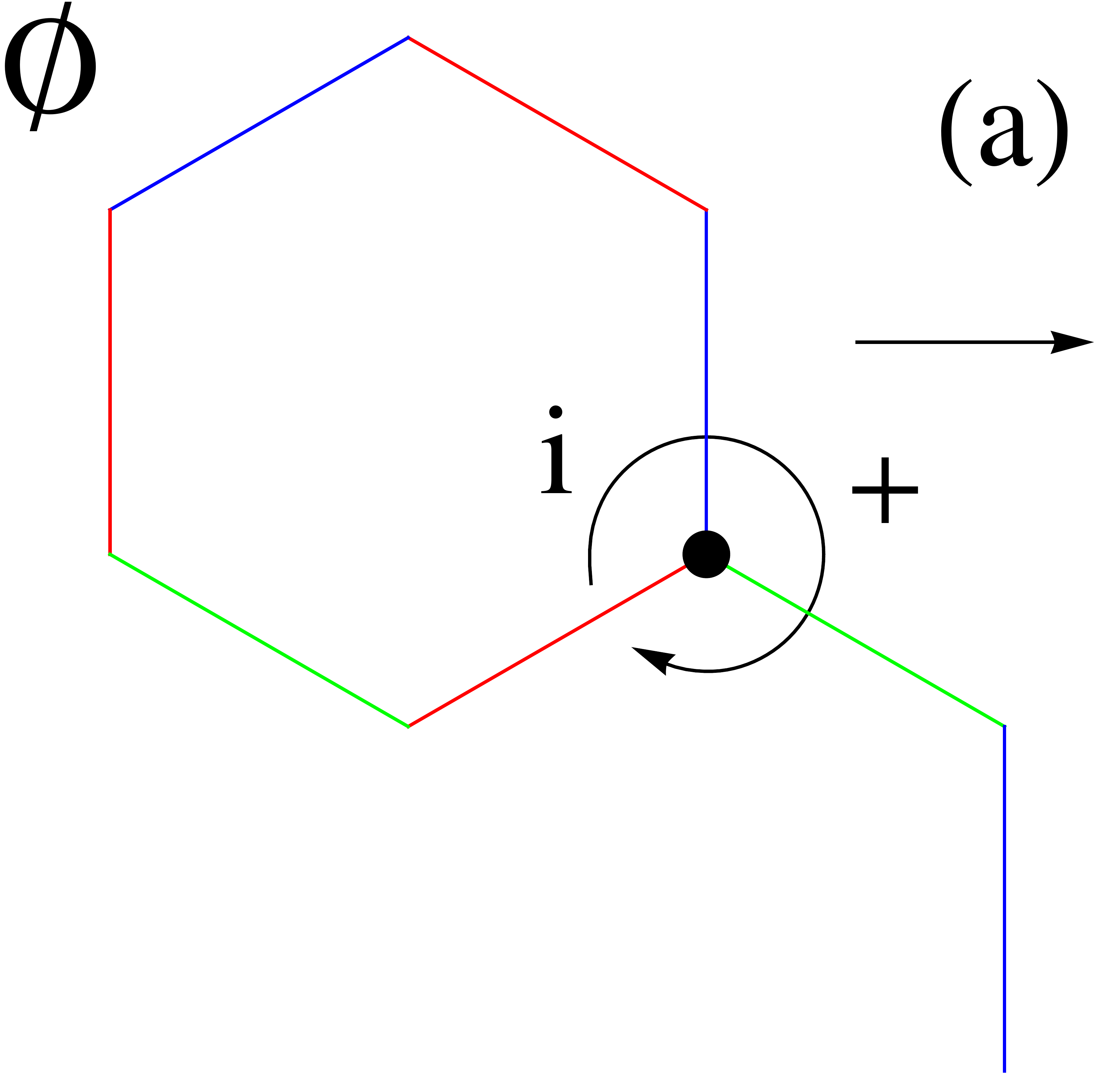,width=3.0cm,angle=-0} \hspace{1cm}
\psfig{file=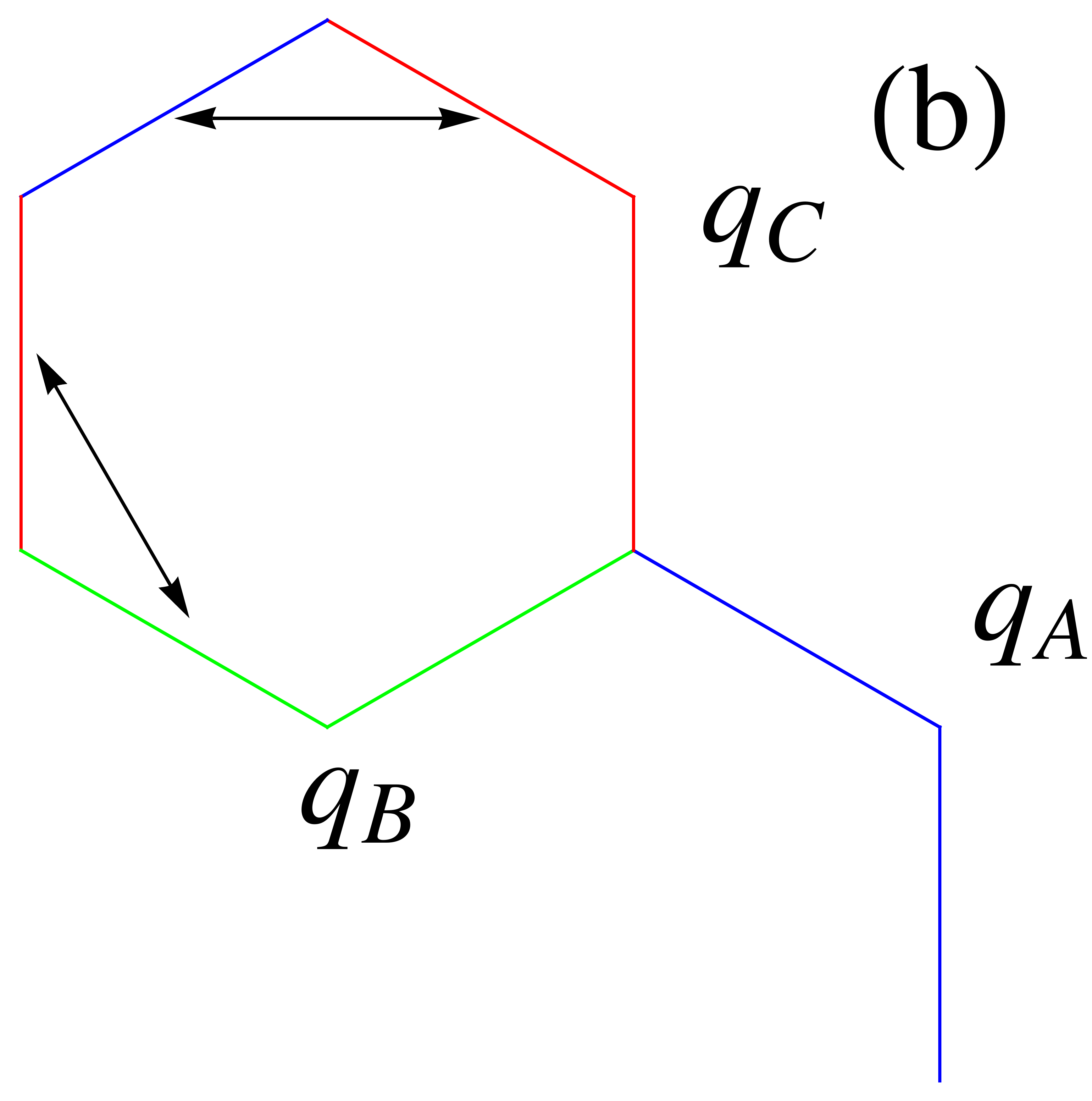,width=3.0cm,angle=-0} \hspace{1cm}
\psfig{file=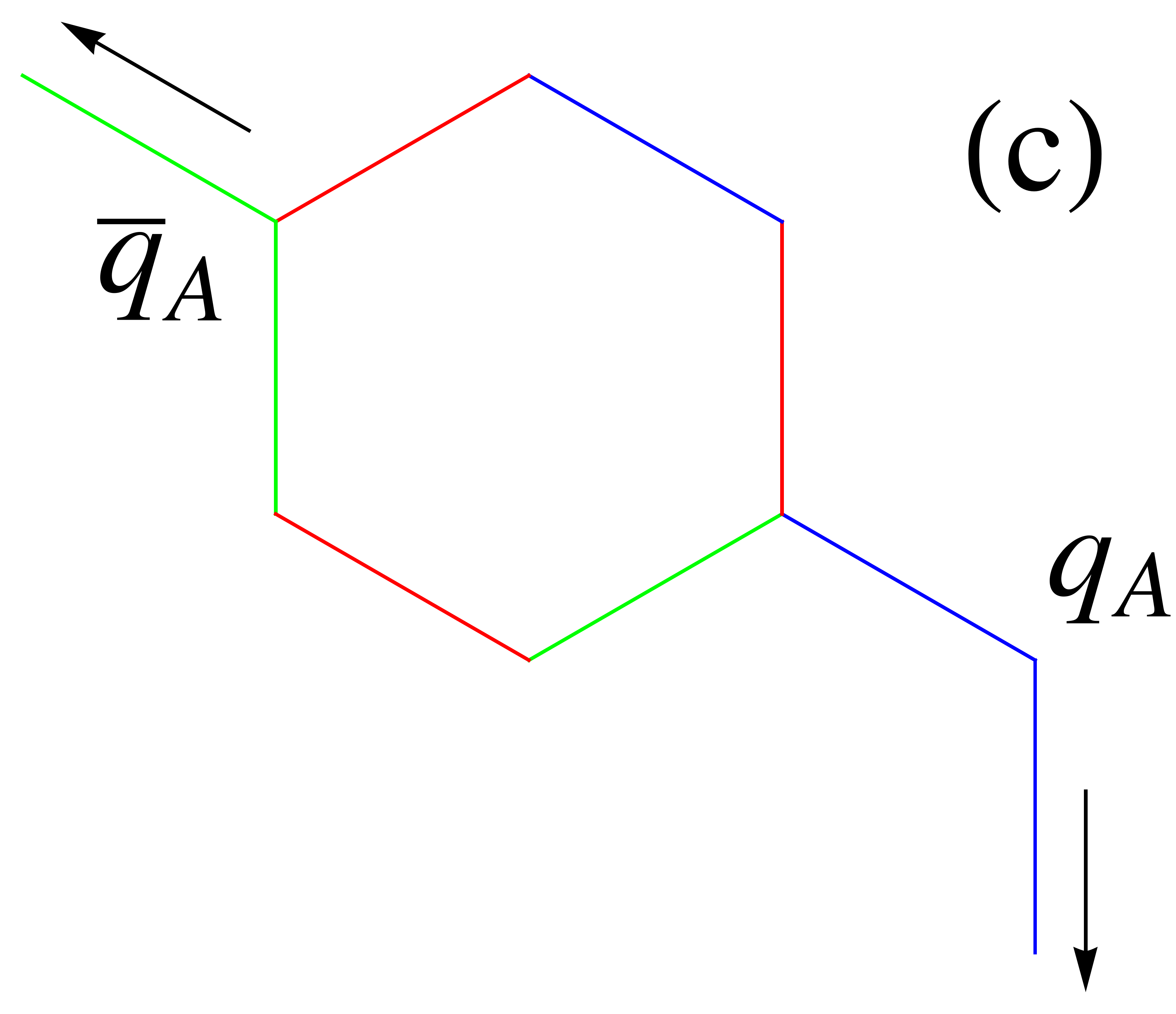,width=3.0cm,angle=-0} \hspace{1cm}
\psfig{file=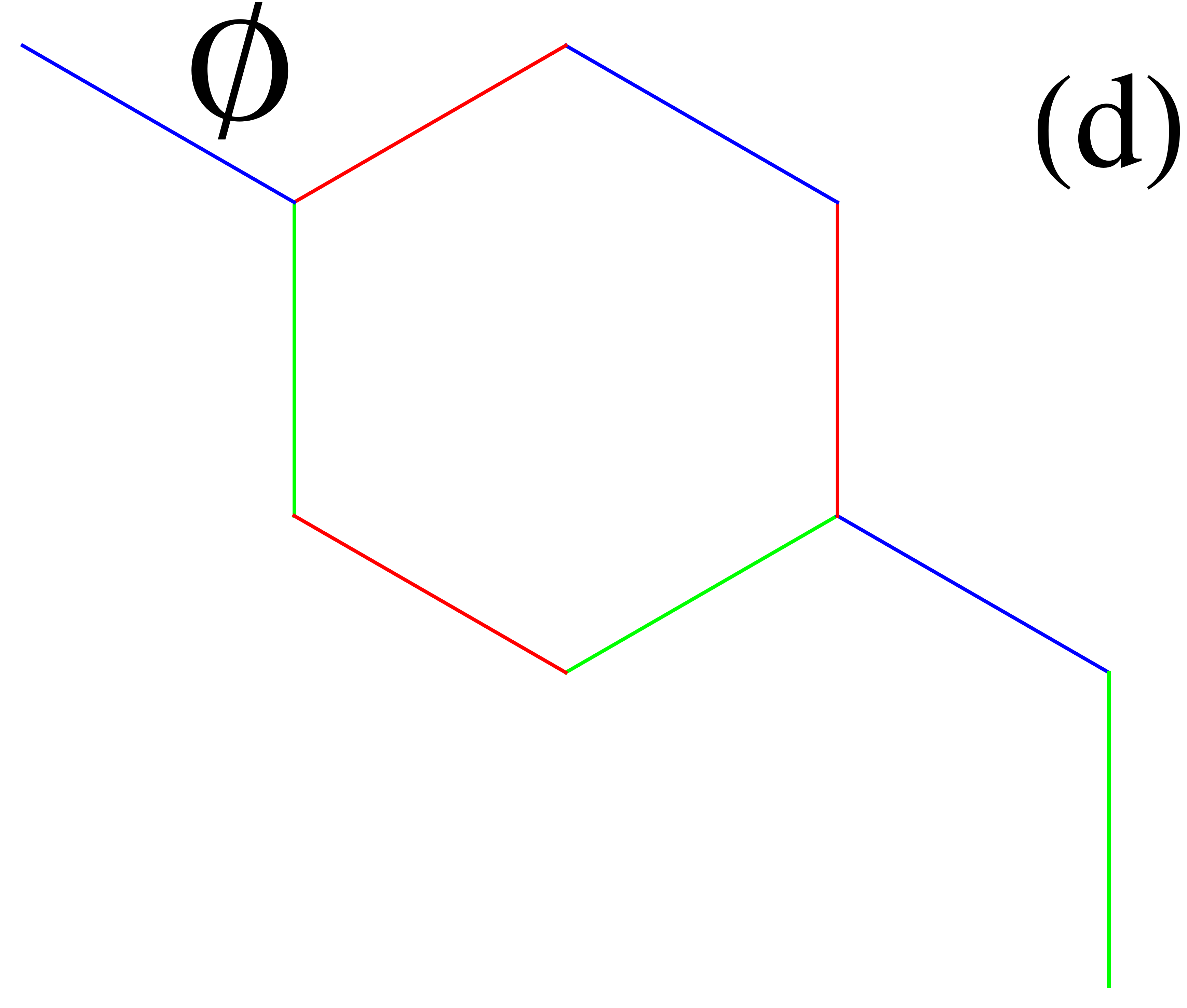,width=3.0cm,angle=-0} \hspace{1cm}
\caption{A way to implement numerically the flip of triple-stranded loops, by using defects: (a) from a perfect three-coloring ($\emptyset$), the three colors of the vertex $i$ marked by a point are exchanged, here clockwise $+$. (b) This generates three defects $q_A$ (BBR), $q_B$ (GGB) $q_C$ (RRG). $q_B$ propagates on a G-R loop, $q_C$ on a R-B loop (double arrows in (b)) until they annihilate $q_B+q_C \rightarrow \bar{q}_A$ (c). The last two defects propagate on a G-B loop and recombine in a perfect three-coloring $q_A+\bar{q}_A \rightarrow \emptyset$, resulting in the overall flip of a triple-stranded loop (d).}
\label{algodetails}
\end{figure}

The other process that we exploit here is to create a triplet of defects $q_A$, $q_B$ and $q_C$ (or $\bar{q}_A$, $\bar{q}_B$ and
$\bar{q_C}$) by exchanging the three colors at a given vertex, by a
clockwise or anti-clockwise rotation (Fig.~\ref{algodetails}(a)). We propagate
them away from the vertex on their respective two color loops until
two of them meet (Fig.~\ref{algodetails}(b)). When they meet, they transform onto the conjugate
defect of the third defect (Fig.~\ref{algodetails}(c)). It is then sufficient to further propagate
the third defect until it annihilates with its conjugate (Fig.~\ref{algodetails}(d)). Once the
first vertex and the orientation of the color exchange are chosen, the
entire process is completely determined. The process $I_i^+$, where
$i$ refers to a chosen vertex and $+$ to the orientation, is
summarized as follows,
\begin{eqnarray}
&& \emptyset \rightarrow q_A+q_B+q_C \\
&& q_B+q_C \rightarrow \bar{q}_A \label{meet}\\
&& q_A+\bar{q}_A \rightarrow \emptyset \label{laststep}
\end{eqnarray}
and the conjugate process, $I_i^-$, consists of exchanging the three
colors by the opposite rotation, $\emptyset \rightarrow
\bar{q}_A+\bar{q}_B+\bar{q}_C$. Note that the process (\ref{meet}) may
occur with any pair. The ``time-reversal'' process is not identical in
general, in particular $I_i^-$ is not the reverse process: $I_i^-I_i^+
\neq Id$. This is because in the last step [Eq.~\ref{laststep}], the
propagation of the $q_A$ defect may cross the first two segments. In
this case, this leads to a local reorganisation of these segments. The
process $I_i^-$ starts from the same vertex but does not find the same
segments, since they have been reorganized. A new state different from
the original state is generated.  Such absence of microreversibility
breaks the detailed balance and the Monte-Carlo algorithm would
fail. We have therefore restricted explicitly the motion to those
flips which satisfy $I_i^+I_i^-=Id$ (typically 40-50\% depending on
the size considered).

The Monte-Carlo algorithm works as follows: at each step we choose
randomly whether to flip a loop or a ``stranded'' loop. If it is a
``stranded'' loop, we choose randomly a vertex and an orientation and
create the triplet of defects that we propagate until they annihilate
according to the description given above. We accept only the moves
that are reversible.

We have checked on small lattices that we can now reach all states by
successive application of these moves: for this we let the Monte-Carlo
algorithm run until all distinct states are generated, the number of
which we know from section \ref{enumeration}.  We note that
restricting it to short processes given in Fig.~\ref{overlaps} (top),
\textit{i.e.} with two strands making a small hexagon, does change the
parity but does not generate all states, so we have to include
both. We therefore expect the present algorithm to be ergodic.

\subsection{Results}
\label{Results}

We have prepared  samples of states by using the algorithm described
above. In this dynamics, the parity is no longer conserved and the
parity-parity autocorrelation function decays in less than a
Monte-Carlo sweep ($N$ Monte-Carlo attempts).  
\begin{figure}[h]
\vspace{-.3cm}
\psfig{file=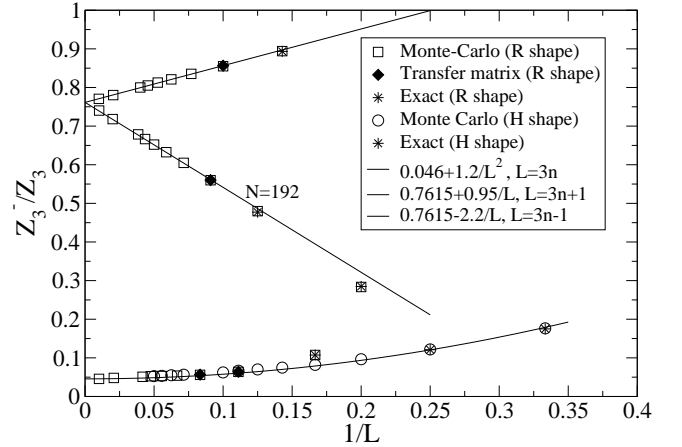,width=9.5cm,angle=-0} \vspace{-.8cm} 
\caption{Fraction of odd states as a function of $1/L$ at fixed aspect ratio $r=1$, depending on mod($L,3$), obtained by Monte-Carlo algorithm (section \ref{ergodic}), exact enumeration up to $L=8$ (section \ref{enumeration}) and transfer matrix up to $L=12$ (section \ref{oddeventm}).}
\label{oddfraction}
\end{figure}
\begin{figure}[h]
\vspace{-.3cm}
\psfig{file=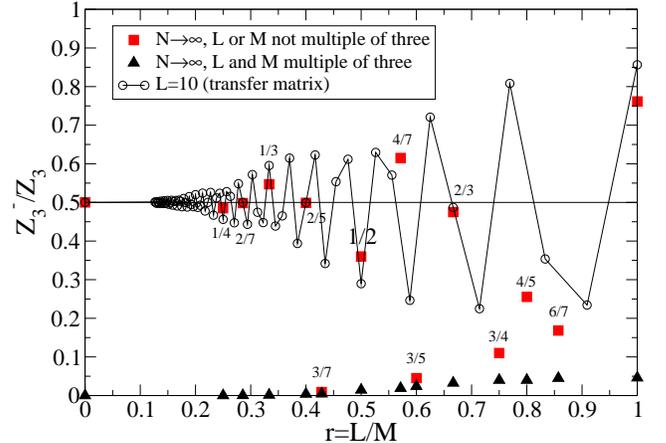,width=9.5cm,angle=-0} \vspace{-.8cm} 
\caption{Fraction of odd states extrapolated in the thermodynamic limit (squares and triangles), as a function of the aspect ratio $r=L/M$, from Monte-Carlo. It depends on whether $L$ or $M$ are multiple of three. The results from transfer matrix at fixed $L=10$ (circles) are shown for comparison.}
\label{oddfractionaspect}
\end{figure}

The first measurement is the fraction of odd states, $Z_3^-/Z_3$ in
the sample of states. The result for $L=M$ is given in
Fig.~\ref{oddfraction}. For small clusters, we recover the exact
results obtained by enumeration (table \ref{tableKT}) and transfer
matrix calculation (table \ref{tabletm1}). In the thermodynamic limit,
we clearly see that the results extrapolate to a finite density, thus
confirming that even and odd classes have the same entropy. For $L$ a
multiple of three and for the rhombus shape or for all clusters with
the hexagonal shape, the fraction is small, $0.046$ in the
thermodynamic limit (bottom points in Fig.~\ref{oddfraction}). Both
are special in that they can accomodate the $\sqrt{3} \times \sqrt{3}$
state without domain walls. When $L=3n\pm 1$ is not a multiple of
three, there are two distinct results (upper points in
Fig.~\ref{oddfraction}), which both converge linearly to a large
fraction of odd states, $0.7615$. A majority of states is, therefore, missed by standard loop Monte-Carlo algorithms in this case.

We have done the same
calculation for different aspect ratios $r=L/M$ and the results of
similar extrapolations are summarized in
Fig.~\ref{oddfractionaspect}. When $L$ and $M$ are multiples of three, the
fraction remains small and smoothly interpolates between 0.046 for
$r=1$ and 0 for $r=0$ (triangles in Fig.~\ref{oddfractionaspect}). When $L$ or $M$ is not a multiple of three,
the fraction varies rapidly as a function of the aspect ratio (squares in Fig.~\ref{oddfractionaspect}). For
comparison, we also give the result of the transfer matrix at finite
$L=10$, where we have a fast-oscillating part.  There may be,
therefore, finite aspect ratios where the Z$_2$ symmetry found at
$r=0$ holds as well. For instance, for $r=2/5$ or $2/7$, we cannot
numerically distinguish the result from $Z_3^-/Z_3=1/2$.

\begin{figure}[h]
\vspace{-.3cm}
\psfig{file=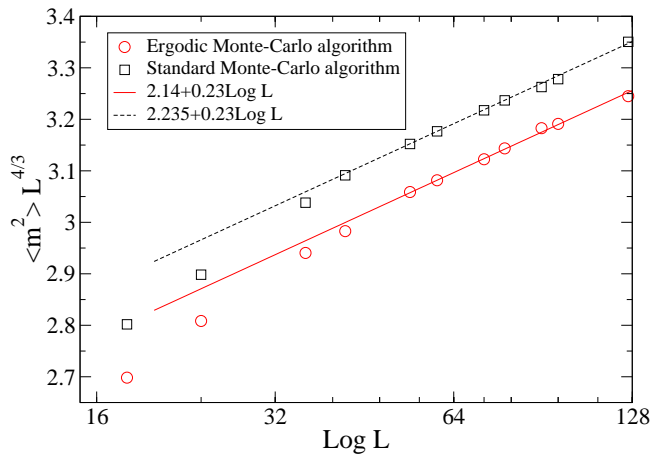,width=9.5cm,angle=-0} \hspace{1cm}
\vspace{-.8cm}
\caption{Order-parameter scaling of the $\sqrt{3} \times \sqrt{3}$ order, for both the standard loop Monte-Carlo algorithm (squares) and the present ergodic algorithm (circles) with periodic boundary conditions and $L=M=3n$. The systematic error when using the standard algorithm is 4.4\%, which reflects the fraction of odd states given in Fig.~\ref{oddfraction}.}
\label{orderparameter}
\end{figure}

The second measurement is the order-parameter of the $\sqrt{3} \times
\sqrt{3}$ state, which was defined in section \ref{chirality}. $L$ has
to be a multiple of three in this case and we know that the fraction
of odd states is small, so we expect that the nonergodic algorithm has a small error. Nevertheless, the issue may be
relevant since it was predicted that the Heisenberg model can be
mapped onto the three-coloring problem with a finite interaction strength of a
few percents, favoring long-range order of the $\sqrt{3} \times
\sqrt{3}$ state.\cite{Chern} The order-parameter we find from the
ergodic algorithm follows the same power-law with $<m^2> \sim L^{-\eta}$ ($\eta=4/3$) as before, but with a shift in the logarithmic correction (Fig.~\ref{orderparameter}). The result is in fact smaller by 4.4\% (independent of $L$ at the numerical precision), reflecting that the small class of odd states
has a much lower order-parameter. Moreover, since the estimate is lower, one needs
a larger interaction to fit the Monte-Carlo data of the Heisenberg
model,\cite{Chern} that therefore strengthens the order predicted
there.

\section{Conclusion}

We have shown that the total chirality of a three-coloring can be an
odd or an even number and defines two classes. This parity is
conserved by the loop dynamics because the lengths of the loops are
even, when periodic boundary conditions are enforced.  Previously-used
loop Monte-Carlo algorithms are trapped in one sector and this
explains the nonergodicity previously noted. This is true for the
torus and Klein bottle, but not for the cylinder or plane where the loops can be of odd lengths.

The odd and even classes generically have the same entropy in the
thermodynamic limit. An exception is the infinitely-long strip of
hexagonal lattice with finite circumference $L$ where the entropy per
site of the odd states is smaller than that of the even states by a
universal correction, $8\pi/(\sqrt{3}L^2)$, when $L$ is large and a multiple of
three (Eq.~\ref{entropyodd}). When $L$ is not a multiple of three,
however, not only the entropies are identical, but also the number of
odd and even states in the thermodynamic limit, so that the system has
an infinite-temperature Z$_2$ symmetry.  For general aspect ratio other than zero, this
Z$_2$ symmetry is absent, but we assume it may exist at special points.

We have argued that ``stranded'' loops make the Monte-Carlo algorithm
ergodic and allow to compute the fraction of odd states and
order-parameters. 
By constrast, the standard loop algorithms that conserve parity lead to a systematic error, that can be large when $L$ is not a multiple of three.
When $L$ is a multiple of three, however, the error on the order-parameter is of a few \%, 
because, in this case, the odd class (although still extensive) is
small.

\acknowledgements

I would like to thank J.-C. Angl\`es d'Auriac, L.~Jaubert and G.~Misguich for discussions and especially A. Ralko for his interest and participation.

\appendix

\section{Enumeration of dimer coverings on hexagonal lattices}
\label{appA}

The method of enumeration of dimer coverings is known,\cite{Kasteleyn,Fisher} but the actual numbers for finite-size hexagonal lattices are known only in a few cases.\cite{Elser,Schwandt}
We recall the method for completeness and for the discussion of appendix \ref{invariantdimer} and  \ref{alternatives}.

\subsection{Hexagonal shape with open boundary conditions}

We consider the dimer problem on the hexagonal lattice shown in Fig.~\ref{Hexlatt} with open boundary conditions. In this case, we recall that\cite{Kasteleyn,Fisher}
\begin{equation}
Z=\sum_D 1 = |\mbox{Pf} K|
\end{equation}
where $\mbox{Pf}$  is the Pfaffian of the Kasteleyn matrix $K$, defined by $K_{ij}=\pm 1$ if $i$ and $j$ are neighbours with the sign definition given in Fig.~\ref{Hexlatt}, $K_{ji}=-K_{ij}$.
By definition, the Pfaffian of an antisymmetric matrix $K$ of even size $N$ is
\begin{equation}
\mbox{Pf} K = \sum_P \sigma(P) K_{i_1i_2}K_{i_3 i_4} \cdots K_{i_{N-1},i_N}
\end{equation}
with the restriction that $i_1<i_2$, $i_3<i_4$ \textit{etc.} and $i_1<i_3<i_5
\dots$. The sites $i_1,i_2,\dots$ are obtained by a permutation $P$
of $1,2,...,N$ and $\sigma(P)$ is its signature
($\sigma(P)=(-1)^{\#transpositions}=(-1)^{n_e}$ where $n_e$ is the number of even
cycles). By an appropriate choice of signs of the matrix elements
$K_{ij}$ one can compensate the negative sign of the permutation and
obtain a sum over all configurations with weight 1, \textit{i.e.} enumerate all
states.\cite{Kasteleyn,Fisher} There are $N/2$ terms in the product of matrix
elements. 
Kasteleyn showed that for the honeycomb lattice one has to choose a $-$ sign for the product of all signed edges around each hexagon, which is obtained by the choice given in Fig.~\ref{Hexlatt}. 
For a bipartite lattice, the matrix $K$ has the form,
\begin{equation}
K= \begin{pmatrix} 0 & D \\
-D^T & 0
\end{pmatrix} 
\end{equation}
and $\mbox{Pf} K = \pm \mbox{det} D$. It is therefore sufficient to
compute a single determinant of a matrix of size $N/2 \times N/2$. One
can see it directly from the definition of the determinant, $\mbox{det} D = \sum_P \sigma(P) D_{1i_1}D_{2 i_2} \cdots D_{N/2,i_{N/2}}$. The first indices $1,\dots,N/2$ denote the ``black'' sites and the second indices any permutation $P$ of the ``white'' sites. In this way every ``black'' site is paired to a ``white'' site, but we do not want long distance pairing so the matrix elements have to be zero except for the nearest neighbours. In this case, every site that is paired appears only once, and so every term in the sum corresponds to a dimer configuration. It is then important to fix the sign of the summand, so that each configuration is counted with the plus sign.
Let us start with the reference
configuration shown in Fig.~\ref{Hexlatt}, sometimes called the ``empty room'' by analogy with the problem of 
plane partitions.\cite{Elser} In this configuration, the dimers are alternating in
onion rings around the center. In terms of
permutation, it is the identity, which
consists of pairing a black  site $i$ with the white site $i$. The product of $D_{ii}$ is positive, so that the reference
configuration is counted with a plus sign (an alternative sign
configuration is to take all arrows from black to white sites, in this
case $K_{i,j}=+1$). Any cyclic motion around an hexagon corresponds to
a cycle of odd length (half the length of the loop) that has a positive signature $(-1)^{l/2-1}=+1$, and a positive product of
edges. Since any state can be reached through a sequence of cyclic
motion, every dimer state is counted with a positive sign, and
therefore \begin{equation} Z= \mbox{det} D.\end{equation} The determinant of the $N/2 \times N/2$ matrix $D$ is computed numerically and the results are given in table
I. In fact, these numbers are known: the problem is exactly that of plane partitions\cite{Elser} and the numbers of them for a $a \times b \times c$ hexagonal shape are the MacMahon numbers:
\begin{equation}
Z = \prod_{i=1}^a \prod_{j=1}^b \prod_{k=1}^c \frac{i+j+k-1}{i+j+k-2} 
\label{MacMahon}
\end{equation}
matching those in table \ref{hexopen} with $a=b=c=L$.
\begin{table}
\begin{tabular}{c|l}
L & $Z$ \\
\hline 
\hline
 1 & 2 \\
 2 & 20 \\
 3 & 980 \\
 4 & 232848 \\
 5 & 267227532 \\
 6 & 1478619421136 \\
 7 & 39405996318420160 \\
 8 & 5055160684040254910720 \\
 9 & 3120344782196754906063540800 \\
 10 & 9265037718181937012241727284450000 \\
 11 & 132307448895406086706107959899799334375000 \\
\hline
\end{tabular}
\caption{Numbers of dimer coverings with open boundary conditions, computed from the determinant or from the MacMahon formula (Eq.~\ref{MacMahon}) with $a=b=c=L$.}
\label{hexopen}
\end{table}

\subsection{Hexagonal shape with toroidal periodic boundary conditions}
\label{hexper}

It is not possible to write the partition function as a single
Pfaffian with periodic boundary conditions (the graph is not planar)
since not all states can be reached with cyclic motion of hexagons. We need
 a linear combination of four Pfaffians.\cite{Kasteleyn} The
reference dimer state (Fig.~\ref{Hexlatt}) corresponds to the identity in terms of 
permutations and has a signature $+1$ and a positive product of
oriented edges, for all weights $(u_1,u_2,u_3)$ on the boundaries (for $x,y,z$ orientations). It has $L$
dimers along a line that cuts $3L$ edges (one is shown by a dashed
line in Fig.~\ref{Hexlatt}) and there are three such lines along the
three principal direction. Depending on the parity of $L$, the
reference state has, therefore, either an (odd,odd,odd) or
(even,even,even) number of dimers along the three directions: for odd $L$, all the states in the
(odd,odd,odd) sector are counted with a plus sign. Now there are
states that differ by a loop winding accross the boundaries, such a
loop can change the parity of two sectors (so that the sum
remains of the same parity as that of $L$), say 1 and 2, so the state
is in a (odd,odd,even) sector for a reference state in a
(even,even,even) sector. Since the winding loop also corresponds to an
odd cycle, the signature of the new state does not change.
 
For this reason, the number of dimer coverings is 
\begin{eqnarray}
Z= \frac{1}{2} [  \mbox{det} D(1,1,1) - \mbox{det} D(1,-1,-1) + \nonumber \\
 - (-1)^{L} \mbox{det} D(-1,-1,1) - (-1)^L \mbox{det} D(-1,1,-1)  ] 
\end{eqnarray}
where $D(u_1,u_2,u_3)$ is a Kasteleyn matrix with signs ($u_1=\pm 1,u_2=\pm 1,u_3=\pm 1$ and the constraint $u_1u_2u_3=1$) on the periodic boundaries, according to the three edge directions. The determinant of each $N/2 \times N/2$ matrix $D$ is computed numerically and the results are given in Table \ref{hexperiodic}. These numbers match those obtained by construction, in Table \ref{table1}. 

\begin{table}
\begin{tabular}{c|l}
L & $Z$ \\
\hline 
\hline
 1 & 6 \\
 2 & 120 \\
 3 & 15162 \\
 4 & 13219200 \\
 5 & 80478777786 \\
 6 & 3417194853335640 \\
 7 & 1010200119482131248768 \\
 8 & 2077088937091136948273774592 \\
 9 & 29688796156479320775456569461994826 \\
 10 & 2949240953029338499089605475162868134375000 \\
\hline
\end{tabular}
\caption{Numbers of dimer coverings $Z$ for the hexagonal shape with periodic boundary conditions, computed with four determinants (as in table \ref{table1} for $L\leq 4$). Some of these numbers were known.\cite{Schwandt}}
\label{hexperiodic}
\end{table}

\subsection{Rhombus shape with toroidal periodic boundary conditions}
\label{Rhombtorus}
Here the shape of the cluster is that of a rhombus (Fig.~\ref{Honlatt}). Without periodic boundary conditions, there is a single configuration, consisting of putting all dimers in the same direction, compatible with the boundaries (this is the identity in terms of permutation 1-1, 2-2 \textit{etc.}). With periodic boundary conditions again the graph is nonplanar (with the exception of the $L=2$ cluster) and four Pfaffians are needed. The result is
\begin{eqnarray}
Z= \frac{1}{2} | (-1)^{L+1} \mbox{det} D(1,1) + \mbox{det} D(-1,1) + \nonumber \\ + \mbox{det} D(1,-1) +(-1)^L \mbox{det} D(-1,-1) | 
\end{eqnarray}
where $D(u_1,u_2)$ is a Kasteleyn matrix with additional signs $(u_1=\pm 1,u_2=\pm 1)$ accross the two $(x,y)$ boundaries. The number of dimer configurations is given in table \ref{rhombusperiodic}, in agreement with table \ref{table1}.    

\begin{table}
\begin{tabular}{c|l}
L & $Z$ \\
\hline 
\hline
 2 & 9 \\
 3 & 42 \\
 4 & 417 \\
 5 & 7623 \\
 6 & 263640 \\
 7 & 17886144 \\
 8 & 2249215617 \\
 9 & 547003370634 \\
 10 & 255635055079809 \\
 11 & 223497249280847919 \\
 12 & 379028233842678000000 \\
 13 & 1225114320423161720823183 \\
 14 & 7452791939816339215874217984 \\
 15 & 87934912263192096558472630935552 \\
 16 & 1969541555284024563005131046158940673 \\
\hline
\end{tabular}
\caption{Numbers of dimer coverings $Z$ for the rhombus shape with periodic boundary conditions (as in table \ref{table1} for $L \leq 8$).}
\label{rhombusperiodic}
\end{table}

\section{A general invariant in the dynamics of dimer coverings?}
\label{invariantdimer}
Each dimer covering has a possible invariant under dimer exchange along the loops,
\begin{equation}
I_d= \mbox{det} [n_{ij} D_{ij}(u_1,u_2)],
\end{equation}
where $n_{ij}=1$ if there is a dimer on bond $ij$ and 0 otherwise, and
$D_{ij}$ is the Kasteleyn matrix with $u_1$ and $u_2$ weights on the
boundaries (on nonbipartite lattices, one has to replace the
determinant by a Pfaffian, see appendix \ref{appA}). We first discuss the case $u_1=u_2=1$.  On a planar graph, $I_d=1$ 
for all dimer coverings.\cite{Kasteleyn} On a nonplanar graph, $I_d=\pm
1$ is in general not an invariant but it may remain invariant
in a class of states. For instance in the problem of section
\ref{Rhombtorus}, $I_d=1$ for all states in the (even,even)
class. This is ensured by the $\pi$ flux condition on each plaquette:
a dimer permutation of a hexagon is a cycle of odd length so it has
$\sigma(P)=1$ and a product of $D_{ij}>0$. Since all closed loops have
length $4n+2$ on the honeycomb lattice, all cycles are odd and the
invariant is therefore conserved. This is no longer true on a general
lattice, as shown on the cubic lattice, where the flip of a square
plaquette conserves $I_d$ (square cycles $(12) \rightarrow (21)$ have an
odd signature, and the product of signed edges is odd, so that
$I_d=1$).\cite{Freedman} But there are longer closed loops that
correspond to odd cycles and odd number of negative edges, so that in
this case the invariant is restricted to moving the smallest
loops.\cite{Freedman}

On the honeycomb lattice with periodic boundary conditions of the torus geometry, this remains true for \textit{all} closed loops and
$I_d$ differs only in different winding sectors.  For instance, in the rhombus case, $I_d=-1$
whenever the state belongs to a sector with $N^x$ or $N^y$ odd, or
both odd. For a general $u_1, u_2$, the considerations explained in appendix \ref{appA} lead to
\begin{equation} I_{d}(u_1,u_2)=(-1)^{N^x+N^y+N^xN^y} u_1^{N^x} u_2^{N^y}. \label{eqI} \end{equation}

\section{Alternative descriptions of the parity invariant}
\label{alternatives}

\subsection{Product of signatures of permutations}

We have seen that a three-coloring of the edges can be seen as three
non-overlapping dimer coverings. Each dimer covering can be written as
a permutation of the lattice sites, characterized by an invariant
$I_d=\pm 1$ (see appendix \ref{invariantdimer}).  We may thus define the
quantity
\begin{equation}
I_3 \equiv I_d(A) I_d(B) I_d(C) = \pm 1,
\end{equation}
where $I_d(i)$ corresponds to the invariant $I_d$ of each of the three dimer
coverings of color $i$. We show that $I_3$ is actually conserved by
the motion of winding loops: when we flip an A-B winding loop, it corresponds to a cycle $C$ in the permutation of dimers 
$A$ and the cycle $C^{-1}$ in the permutation of dimers $B$. The product of these two cycles corresponds to 
exchanging $I_d(A)$ and $I_d(B)$, so that $I_3$ is constant.  It turns out
that $I_3$ equals to -1 in the odd sector and to +1 in the even
sector (hence the same notation). Specifying to the rhombus shape with torus geometry, we have (see Eq.~\ref{eqI}) 
\begin{equation} I_d(i)=(-1)^{N^x_i+N^y_i+N^x_iN^y_i} u_1^{N^x_i} u_2^{N^y_i}. \end{equation}  
Noticing that
$\sum_{i=1}^3 N^{\alpha}_i=n$ is a constant that depends only on $L$, which we discard, we obtain
\begin{equation}
I_3 = (-1)^{\sum_{i=1}^3 N_i^x N_i^y}
\end{equation}
Note that what is true for a pair $x,y$ of lattice directions is also true for the other pairs by symmetry. So that $I_3$ can be symmetrized and simplified,
\begin{equation}
I_3 = (-1)^{\sum_{i=1}^3 N_i^x N_i^y+N_i^y N_i^z+N_i^z N_i^x} = (-1)^{\frac{1}{2} \sum_{i \alpha} (N_i^{\alpha})^2}.  
\end{equation}
The argument $\frac{1}{2} \sum_{i \alpha} (N_i^{\alpha})^2$ is
an even or odd integer. In this form, it is transparent that $I_3$ is invariant by all
nonwinding loops, but is it remarkable that it remains invariant under
winding loops as well.

\subsection{Other related colorings}

Fisk has studied a series of colorings and invariants.\cite{Fisk} He
showed that a 4-coloring of the sites of the triangular lattice
induces a 3-coloring of the edges, which in turn induces a ``Heawood''
coloring of the sites and a local coloring. On a regular triangular
lattice with open boundary conditions, we have the result that if $Z$
is the number of 3-colorings, $4Z$ is the number of
4-colorings.\cite{Baxter} This is no longer true on a triangular
lattice with periodic boundary conditions: starting from a valid
4-coloring of the sites, it is always possible to find an edge
coloring.\cite{Fisk} For completeness, we recall that the mapping
associates to two neighbors of the triangular lattice a color of that
bond, according to $(12)$ or $(34)$ gives color $1$ for the edge,
$(13)$ or $(24)$ gives 2, and $(14)$ or $(23)$ gives 3. But the
converse is not true: there are more three-colorings of the edges than
4-colorings (divided by four) of the sites. Fisk also defines a
"Heawood coloring'' which is what is called the chirality here, and a
``local'' coloring which is an anti-domain wall separating sites of
identical chirality. 
The number
of such ``singular edges'' can be odd or even and is actually directly
related to the number of vertices with positive
chiralities.\cite{Fisk} So that the odd/even invariant is also that of
the number of anti-domain walls. No other simple invariants are known.\cite{Fisk}

\subsection{Odd/even invariant of graphs}
 
The parity of $m$ turns out to be also related to the odd-even invariant
recently introduced for colorings and graphs.\cite{Eager} Let us make the
connection explicit: given an orientation of the lattice, Eager and
Lawrence count the number of oriented bonds $i \rightarrow j$ (where
$i$ and $j$ are edges in our case) with $\sigma_j>\sigma_i$.  The
parity of this number is the ``odd-even invariant''.\cite{Eager} To
make the connection with the parity of $m$ explicit, take each
vertex, call the edges 1-2-3 clockwise and orient the bonds from edge
1 to edge 2, from edge 2 to edge 3, and from edge 1 to edge 3.  All
configurations with + (resp. -) chirality have one or three edges
satisfying the rule above, \textit{i.e.} an odd number $p$ of edges (resp. zero
or two, \textit{i.e.} an even number $n$ of edges). The total number of bonds
satisfying the rule is $n_e= p N_+ + n N_- $ where $N_{\pm}$ are the numbers of vertices with $\pm 1/2$ chirality (note that there are three bonds for each vertex). When $N_+$ is even,
$N_-$ is even (because $N_++N_-=2N/3$ is even) so $n_e$ is even. When
$N_+$ is odd, $pN_+$ is odd and $nN_-$ is even, so $n_e$ is
odd. Therefore a state with  $I_3=\pm 1$ invariant is an even/odd
state in the language of Ref.~\onlinecite{Eager}.

\section{Dimer and three-colorings on the Klein bottle}
\label{SM}

\begin{figure}[h]
\psfig{file=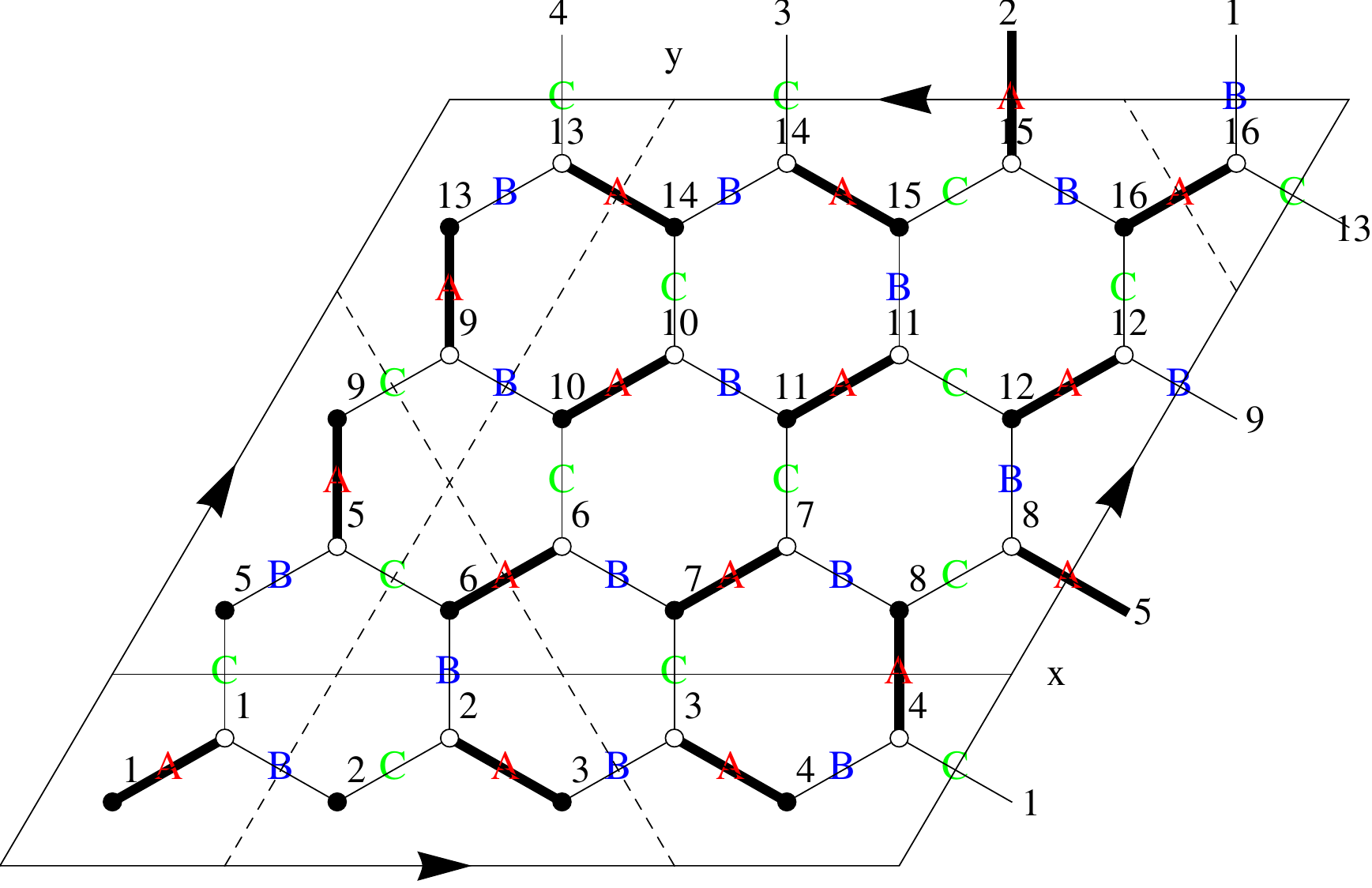,width=8.0cm,angle=-0}
\caption{Hexagonal lattice with periodic boundary conditions of Klein bottle geometry, see the arrows on the sides. A dimer and color configurations are shown. Here a rhombus (R) shape is chosen with $L=4$ and $N=3L^2=48$ edges. The solid and dashed line show two distinct nonlocal cuts that define topological conserved numbers.}
\label{HonlattSM}
\end{figure}

We consider the dimer and three-coloring problems on an hexagonal lattice with a ``Klein bottle'' geometry of the boundaries. The model is defined on finite-size clusters of size $L, M$ and rhombus shape; see Fig.~\ref{HonlattSM} for an example. While the right boundary is identical to that of the torus, the upper boundary is flipped as in a M\"obius strip, as shown by the arrows.

\subsection{Exhaustive construction}

We have constructed explicitly all the dimer coverings and
three-colorings on the clusters with $L=M$ up to $L=8$.  The numbers
are given in table~\ref{tableK}. On the Klein bottle, the three
directions are nonequivalent and the number of dimer configurations is
therefore no longer a multiple of three in general (but the number of
three-colorings remains a multiple of six). The numbers are different
from those obtained on the torus but remains of the same order of
magnitude.
\begin{table}
\begin{center}
\begin{tabular}{rrrrr}
  \hline \hline
  $N$ & L & T & $Z^K$ & $Z_3^K$    \\
  \hline
  12 & 2 & R & 9 & 24   \\
  27 & 3 & R & 44 & 144  \\
  48 & 4 & R & 425 & 1,824   \\
  75 & 5 & R & 7,751 & 50,496 \\
  108 & 6 & R & 269,200 & 3,250,560  \\
147 & 7 & R & 18,031,040 & 453,925,632  \\ 
192 & 8 & R & 2,283,471,985 & 124,786,807,296  \\ 
  \hline \hline
\end{tabular}
\end{center}
\caption{Numbers of dimer coverings $Z^K$ and three-colorings $Z_3^K$ on the ``Klein bottle'' (rhombus shape, see Fig.~\ref{HonlattSM}). $Z^K$ is also calculated from Pfaffians and $Z_3^K$ from transfer matrix below.}
\label{tableK}
\end{table}

\subsection{Enumeration of dimers by Pfaffian}

The number of dimer coverings on the Klein bottle has been studied in Refs.~\onlinecite{Lu}. It is given by
\begin{equation}
Z= | \mbox{Re} [ \mbox{det} D((-1)^{L+1},\imath) ] | +|  \mbox{Im} [ \mbox{det} D((-1)^L,\imath) ]|  
\end{equation}
where the matrix $D(u_1,u_2)$ is the Kasteleyn matrix deduced from Fig.~\ref{HonlattSM} and $\imath^2=-1$ ensures the correct sign of the (even,even) sector. The determinant is computed numerically in table \ref{rhombusperiodicKlein}, and the numbers match those of table \ref{tableK}.
\begin{table}
\begin{tabular}{c|l}
L & $Z^K$ \\
\hline 
\hline
 2 & 9 \\
 3 & 44 \\
 4 & 425 \\
 5 & 7751 \\
 6 & 269200 \\
 7 & 18031040 \\
 8 & 2283471985 \\
 9 & 554296573020 \\
 10 & 257422540282721 \\
 11 & 226671176777404967 \\
 12 & 382906922419021541632 \\
 13 & 1233881647743136383304247 \\
 14 & 7553274215848727289369432064 \\
 15 & 88700806949845037589354602938368 \\
 16 & 1984263088240036324309600061358282721 \\
\hline
\end{tabular}
\caption{Numbers of dimer coverings $Z^K$ on the Klein bottle (rhombus shape), as in table \ref{tableK} for $L \leq 8$.}
\label{rhombusperiodicKlein}
\end{table}

\subsection{Enumeration of three-colorings  by transfer matrix}

The enumeration of three-colorings by transfer matrix needs a simple modification with respect to the torus case.
For the Klein bottle geometry, the first and the last row have to be
images in a $x \rightarrow -x$ mirror symmetry, so we define an
operator $P$ by $P | \sigma_1 \dots \sigma_{L}\rangle=| \sigma_{L}
\dots \sigma_{1}\rangle$ (with $P^2=1$). We then have
\begin{equation}
Z_3^K = \sum_{\sigma_1} \dots \sum_{\sigma_{M}} T_{\sigma_1 \sigma_2} T_{\sigma_2 \sigma_3} \cdots T_{\sigma_{M} \bar{\sigma_1}}  
=
\mbox{Tr} [ T^{M} P ]
\label{TrK}
\end{equation}
Because of the mirror symmetry, $[T,P]=0$, it is possible to simultaneously diagonalize $T$ and $P$, so that
\begin{equation}
Z_3^K =
\mbox{Tr} [ T^{M} P ]= \sum_{i=1}^{3^L} \Lambda_i^{M} \pi_i
\label{TrK}
\end{equation}
where $\pi_i=\pm 1$ is the parity of eigenvector $i$ under the mirror
symmetry. We note that because of the Perron-Frobenius theorem for
matrices with positive entries, the components of the largest
eigenvector can be chosen to be positive. The eigenvector must be
even, so that the sum $Z_3^K$ is positive, as it should be. Note that
we cannot deduce that $Z_3^K<Z_3$ since the $\Lambda_i$ are
complex. The result of the computation of the number of
three-colorings on finite-size systems is given in table
\ref{tabletm1K}. A consequence of Eq.~\ref{TrK} is that $Z_3^K \sim
\Lambda_1^M \sim Z_3$ in the thermodynamic limit $M\rightarrow
+\infty$, so that the entropies of three-colorings on the Klein bottle
or on the torus are the same.

\subsection{Dynamics}

We have studied the existence of sectors in the dynamics. For this, we
have iterated the loop dynamics on finite-size clusters (up to $L=7$),
starting from a single state and generating classes of
states. Similarly to the torus case, there are winding-number sectors and
Kempe sectors.

On the Klein bottle, there are also nontrivial cycles that define
the winding sectors, but contrary to the torus, they are inequivalent. The
cycle in the $x$ direction is identical to that of the torus, so that
$N_A^x+N_B^x+N_C^x=L$ but the cycles in the $y$ or $z$ directions are
twice longer and change direction at the boundary (see the dashed line in Fig.~\ref{HonlattSM}),
$N_A^{y,z}+N_B^{y,z}+N_C^{y,z}=2M$.

We find one or two Kempe classes depending on $L$ in this case, up to lattice
symmetries (table \ref{tableKK}). We emphasize therefore that the
number of Kempe classes depends explicitly on the geometry of the
boundaries, since clusters with exactly the same number of sites do not
have the same number of Kempe classes (compare tables \ref{tableKK}
and \ref{tableKT}).

Although the surface is nonorientable, we can define the chirality for vertices up
to the upper cut. The parity of $m$ is again conserved and is given in
table \ref{tableKK}. The loop Monte-Carlo is therefore similarly
nonergodic on the Klein bottle, and we use the ``stranded'' version to determine the fraction of odd states.
\begin{table}
\begin{center}
\begin{tabular}{rrr}
  \hline \hline
  $N$ & $n_K$ &  $Z^K_{i=1,\cdots,n_K} (\pm)$  \\
  \hline
  12 & 1 & 24(+) \\
  27 & 2 & 132(-),6(x2)(-) \\
  48 & 2 & 1,056(+), 768(-) \\
  75 & 2 & 36,096(-); 7,200(x2)(+) \\
  108 & 2 & 2,600,448(+); 650,112(-) \\
  147 & 2 & 249,212,928(-); 204,712,704(+) \\ 
  \hline \hline
\end{tabular}
\end{center}
\caption{Number of invariant classes $n_K$ for clusters of size $N$ on the Klein bottle, and number of states in each class $Z^K_{i=1,\cdots,n_K}$ ($\sum_{i=1}^{n_K} n_i Z^K_i=Z^K_3$ is given in table \ref{tableK})}
\label{tableKK}
\end{table}

\subsection{Fraction of odd states from Monte-Carlo}

\begin{figure}[h]
\vspace{-.3cm}
\psfig{file=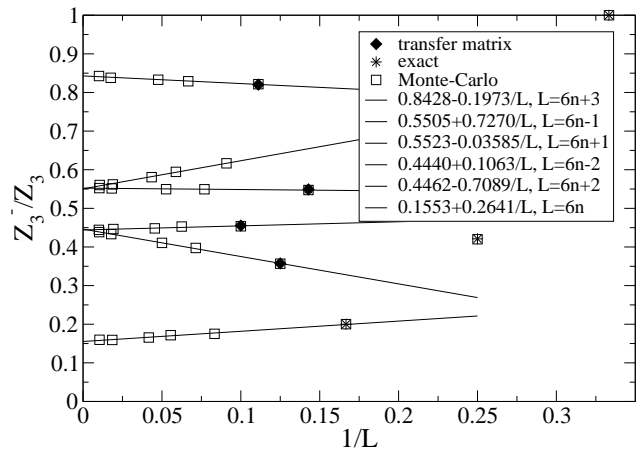,width=9.5cm,angle=-0} \hspace{1cm}
\vspace{-.8cm}
\caption{Fraction of odd states for finite-size clusters of size $L=M$, with the ``Klein bottle'' geometry. There is a mod($L,6$) effect, partially resolved in the thermodynamic limit.}
\label{oddfractionKlein}
\end{figure}

We have computed the fraction of odd states by adding the ``stranded'' loops to the 
Monte-Carlo algorithm. We find that the result extrapolated to the
thermodynamic limit is finite, confirming that odd and even classes
also have the same entropy on the Klein bottle. The fraction now
depends on mod$(L,6)$, but is partially resolved in the thermodynamic
limit (Fig.~\ref{oddfractionKlein}), with pairs $6n\pm 1$, $6n\pm 2$
converging to the same value at the numerical precision, 0.55 and
0.445 respectively.  If we write $p(L)=Z_3^-/Z_3$, we note (without
providing an explanation) that $p(2L)+p(L) \approx 1$, which is specific to the
Klein bottle.

In conclusion, the Klein bottle shares some aspects of the torus with
the same parity invariant. For a Monte-Carlo algorithm to be ergodic,
one has to similarly enrich the allowed motions with ``stranded''
loops. The odd and even classes have also the same entropy, as is
obvious from Fig.~\ref{oddfractionKlein}.

\begin{table}
\begin{center}
\begin{tabular}{rrrrr}
  \hline \hline
  $L$ & $M$  & $Z_3^K$ & $Z_3^{K-}$    \\
\hline
           2 &           2 & 24 & 0 \\
           3 &           2 & 48 & 12 \\
           3 &           3 & 144 & 144 \\
           4 &           2 & 120 & 24 \\
           4 &           3 & 480 & 144 \\
           4 &           4 & 1824 & 768 \\
           5 &           2 & 288 & 132 \\
           5 &           3 & 1536 & 1248 \\
           5 &           4 & 8256 & 5712 \\
           5 &           5 & 50496 & 36096 \\
           6 &           2 & 744 & 240 \\
           6 &           3 & 5184 & 1296 \\
           6 &           4 & 39648 & 12576 \\
           6 &           5 & 350592 & 80640 \\
           6 &           6 & 3250560 & 650112 \\
           7 &           2 & 1872 & 1020 \\
           7 &           3 & 17952 & 13488 \\
           7 &           4 & 194304 & 116976 \\
           7 &           5 & 2471424 & 1523904 \\
           7 &           6 & 32740992 & 18050880 \\
           7 &           7 & 453925632 & 249212928 \\
           8 &           2 & 4824 & 1800 \\
           8 &           3 & 62688 & 15120 \\
           8 &           4 & 956832 & 310080 \\
           8 &           5 & 17236608 & 5086080 \\
           8 &           6 & 323501184 & 104595840 \\
           8 &           7 & 6307763712 & 2120299776 \\
           8 &           8 & 124786807296 & 44607627264 \\
           9 &           2 & 12288 & 7092 \\
           9 &           3 & 219168 & 164016 \\
           9 &           4 & 4704192 & 3012240 \\
           9 &           5 & 121491456 & 86064576 \\
           9 &           6 & 3291241344 & 2339395392 \\
           9 &           7 & 93849672192 & 71017214208 \\
           9 &           8 & 2743266960384 & 2153241150720 \\
           9 &           9 & 82222744421376 & 67451020701696 \\
          10 &           2 & 31560 & 12384 \\
          10 &           3 & 767136 & 189072 \\
          10 &           4 & 23256672 & 8270688 \\
          10 &           5 & 860378112 & 291036480 \\
          10 &           6 & 33588864384 & 13048871040 \\
          10 &           7 & 1385171841024 & 552878592000 \\
          10 &           8 & 58393359785472 & 24928794444288 \\
          10 &           9 & 2514348535314432 & 1105303915459584 \\
          10 &          10 & 109522261290792960 & 49879717261246464 \\
  \hline \hline
\end{tabular}
\end{center}
\caption{Numbers of 3-colorings on the ``Klein bottle'', $Z_3^K$, computed by numerical transfer matrix, for system size $N=3 \times L \times M$ (Fig.~\ref{HonlattSM}). $Z_3^{K-}$ is the number of odd colorings.}
\label{tabletm1K}
\end{table}


\begin{thebibliography}{99}
\bibitem{Newman} M.~E.~J.~Newman and G.~T.~Barkema, \textit{Monte-Carlo Methods in Statistical Physics}, Oxford 
University Press, 1999.
\bibitem{Stillinger} A. Rahman and F. H. Stillinger, J. Chem. Phys. \textbf{57}, 4009 (1972).
\bibitem{Wang} J.-S. Wang, R. H. Swendsen, and R. Kotecky, Phys. Rev. Lett. \textbf{63}, 109 (1989).
\bibitem{Sikora} O. Sikora, N. Shannon, F. Pollmann, K. Penc, and P. Fulde, Phys. Rev. B \textbf{84}, 115129 (2011).
\bibitem{Freedman} M. Freedman, M. B. Hastings, C. Nayak, and X.-L. Qi, Phys. Rev. B \textbf{84}, 245119 (2011). 
\bibitem{Baxter}  R.~J.~Baxter, J. Math. Phys. \textbf{11}, 784 (1970).  
\bibitem{Huse} D. A. Huse and A. D. Rutenberg, Phys. Rev. B \textbf{45}, 7536 (1992), see reference [13] therein.
\bibitem{Mohar} B. Mohar and J. Salas, J. Stat. Mech. P05016 (2010).
\bibitem{Moore} C.~Moore and M.~E.~J.~Newman, J. Stat. Phys. 99, 629-660 (2000).
\bibitem{Castelnovo} C. Castelnovo, C. Chamon, C. Mudry, and P. Pujol,
Phys. Rev. B \textbf{72}, 104405 (2005). 
\bibitem{Cepas} O. C\'epas and A. Ralko, Phys. Rev. B \textbf{84}, 020413 (2011).
\bibitem{Ioffe} L. B. Ioffe, M. Feigel'man, A. Ioselevich, D. Ivanov, M. Troyer, and G. Blatter, Nature \textbf{415}, 503 (2002).
\bibitem{realpbc} Periodic boundary conditions may be achieved by connecting opposite sides or from three-dimensional lithography.   
\bibitem{keplerate} A. M\"{u}ller, M. Luban, C.  Schroeder, R. Modler, P. K\"ogerler, M. Axenovich, J. Schnack, P. Canfield, S. Bud'ko, and N. Harrison, ChemPhysChem 2, 517 (2001).
\bibitem{Fendley} P.~Fendley, J.~E.~Moore and C.~Xu, Phys. Rev. E \textbf{75}, 051120 (2007).
\bibitem{Korshunov} S.~E.~Korshunov, F.~Mila, and K.~Penc, Phys. Rev. B \textbf{85}, 174420 (2012).
\bibitem{methodenum} We construct numerically all the dimer coverings by making
  a tree.  Each node of the tree corresponds to a cluster
  partially-filled with dimers and the generation of a new node
  consists of adding a dimer on neighboring empty sites.  Only the
  final leaves may be proper dimer coverings. The tree is being
  traversed by using a ``depth-first search'' method which adds nodes
  along a branch, before backtracking when a valid or an invalid
  configuration is obtained. It has the advantage to minimize the need
  of memory, by keeping a single branch of the tree, \textit{i.e.}
  $\sim N$ partial states. The states themselves can be kept in memory
  for small sizes or analysed on
  the fly.
\bibitem{multvsdiag}  This method allows to
work with 128-bit integers and compute the exact number of
configurations up to $2^{127} \approx 10^{38}$. Higher exact numbers need using multiprecision packages.
\bibitem{Chandra} P.~Chandra, P.~Coleman, I.~Ritchey, J. Phys. I France \textbf{3}, 591 (1993).  
\bibitem{Chakraborty} B. Chakraborty, D. Das, and J. Kondev, Eur. Phys. J. E \textbf{9}, 227 (2002).
\bibitem{CastelnovoMC} C. Castelnovo, P. Pujol, and C. Chamon, Phys. Rev. B \textbf{69}, 104529 (2004).
\bibitem{Chern} Gia-Wei Chern and R. Moessner,
Phys. Rev. Lett. \textbf{110}, 077201 (2013).
\bibitem{Cepasq} O. C\'epas and B. Canals, Phys. Rev. B \textbf{86}, 024434 (2012); O. C\'epas, Phys. Rev. B \textbf{90}, 064404 (2014). 
\bibitem{technical} To know whether a state is new or not (and avoid
  cycles), we construct a hash table where we insert the new
  states. We keep all states in memory to analyze the collisions in the
  hash table. A compact way to store the states is to use a single bit for the chirality $\pm$ of each vertex (see section
  \ref{chirality} for the definition).  The honeycomb lattice is
  bipartite and, as the largest size considered has $128$ sites
  ($N=192$ edges) we need at most two 64-bit integers (which we hash). In constrast, a
  direct color coding would need two bits per edge, \textit{i.e.}
  three times more memory.  Note that once the color of a first site
  is fixed all other sites are uniquely determined by the chiralities:
  there are three times more color configurations than chirality
  configurations. By symmetry, we can also impose the chirality of the
  first site to be +, so that we need to keep only $Z_3/6$ states in memory. For
  $N=147$, this requires $Z_3*128/(8*6) \sim 1$ Go of memory, but
  $\sim 300$ Go for $N=192$, which is beyond our capabilities.
\bibitem{Belcastro} S.-M.~Belcastro and R. Haas, Discrete Mathematics, \textbf{325}, 77 (2014).
\bibitem{192}
For $N=192$, we cannot keep
all states in memory (see [\onlinecite{technical}]), so we proceed differently to construct the Kempe sectors. For each state in
a given topological sector, we record all the topological sectors it is linked to
by flipping a winding loop.  In a second time, we construct closed
orbits obtained by applying this link matrix iteratively: each sector becomes a
class of topological sectors and we assume in this way that sectors
that have the same topological numbers are always connected, which is true for all the smaller clusters. 
The result is therefore a lower
bound for $n_K$.
\bibitem{Jaubert} L. Jaubert, J. T. Chalker, P. C. W. Holdsworth, and R. Moessner, Phys. Rev. Lett. \textbf{105}, 087201 (2010).
\bibitem{Read} N. Read(unpublished).
\bibitem{Kondev} J. Kondev and C. Henley, Nucl. Phys. \textbf{464}, 540 (1996). 
\bibitem{Kasteleyn} P. W. Kasteleyn, Physica \textbf{27}, 1209 (1961). 
\bibitem{Fisher} M. E. Fisher, Phys. Rev. \textbf{124}, 1664 (1961).
\bibitem{Elser} V. Elser, J. Phys. A: Math. Gen. \textbf{17}, 1509 (1984).
\bibitem{Schwandt} D. Schwandt, Ph.~D thesis, University of Toulouse (2011), https://tel.archives-ouvertes.fr/tel-00783224v1, p. 30; and G. Misguich (private communication).
\bibitem{Fisk} S. Fisk, Adv. in Math. \textbf{25}, 226 (1977).
\bibitem{Eager} R. Eager and J. Lawrence, Eur. J. of Combinatorics, \textbf{50}, 87 (2015).
\bibitem{Lu} W. T. Lu and F. W. Wu, Phys. Lett. A 293, 235 (2002); W. Li and H. Zhang, Physica A 391, 3833 (2012).

\end{thebibliography}
\end{document}